%% file: main.tex
\documentclass[lettersize,journal]{IEEEtran}
\usepackage{amsmath,amsfonts}
\usepackage{algorithmic}
\usepackage{algorithm}
\usepackage{array}
\usepackage{textcomp}
\usepackage{subcaption}
\usepackage{stfloats}
\usepackage{url}
\usepackage{verbatim}
\usepackage{graphicx}
\usepackage{cite}
\usepackage{multirow}
\usepackage{optidef}
\usepackage[colorlinks,citecolor=red,urlcolor=blue,bookmarks=false,hypertexnames=true]{hyperref}

\begin{document}

\title{FMCW-Based Integrated Sensing and Communication System: Design, Implementation, and Experimental Measurements}

\author{Murat Temiz,~\IEEEmembership{~\textit{Member, IEEE,}}
        Colin Horne,
        Matthew A. Ritchie,~\IEEEmembership{~\textit{Senior Member, IEEE,}}
        Christos Masouros,~\IEEEmembership{~\textit{Fellow, IEEE}}
}


\maketitle

\begin{abstract}
This study proposes a radar-centric integrated sensing and communication (ISAC) system utilizing a two-layer modulation scheme for vehicular networks. Frequency-modulated continuous wave (FMCW) chirps are jointly modulated via phase modulation (PM) and index modulation (IM) to transmit data while maintaining sensing as the primary function. To support this, a novel radar signal processing technique is developed to mitigate the impacts of IM and PM on sensing accuracy, alongside a communication receiver architecture designed to successfully demodulate IM and PM data within FMCW chirps. System performance is evaluated through simulations in the 2.4 GHz and 24 GHz bands under Doppler effects, achieving communication throughputs of 25 Mbps and 50 Mbps, respectively. Furthermore, a proof-of-concept hardware implementation is realized, and experimental measurements via a loopback cable are performed to verify the feasibility of the architecture. Finally, it evaluates the fundamental trade-off between communication throughput, sensing accuracy, and out-of-band emission, demonstrating the system's flexibility to dynamically adjust waveform parameters to meet varying operational requirements.
\end{abstract}


\section{Introduction}
\IEEEPARstart{T}{he} increasing amount of radar sensing and wireless communication applications have intensified the congestion within the frequency spectrum, and this congestion is expected to increase significantly, especially in widely employed sub-6-GHz bands. As a result, the concept of integrated sensing and communication (ISAC) emerged as a pivotal technique for future wireless communication and sensing systems \cite{LiuJointRadarCom2020, wei2023integrated}. Since both sensing and communication are essential in vehicular networks, particularly for autonomous driving, automotive radar stands out as one of the most promising application areas for ISAC systems \cite{MoulinISAC2024, LiangPredISAC2025, DokhanchiAdaptiveWaveform2021}.
Traditional sensing and communication systems are built on separate hardware and individual waveforms. This results in considerable hardware, power consumption, and signal processing requirements. Combining sensing and communications on the same hardware will also reduce the size of the system and power consumption, making it an attractive solution to improve energy efficiency. Furthermore, using a unified waveform also efficiently utilizes the frequency spectrum, hence reducing the congestion in the frequency spectrum. {However, it is also important to consider the cost, hardware complexity, and robustness to channel conditions, such as Doppler shift, when designing ISAC waveforms \cite{MaoWaveform2022}.}

\begin{figure}
    \centering
    \includegraphics[width=1\linewidth]{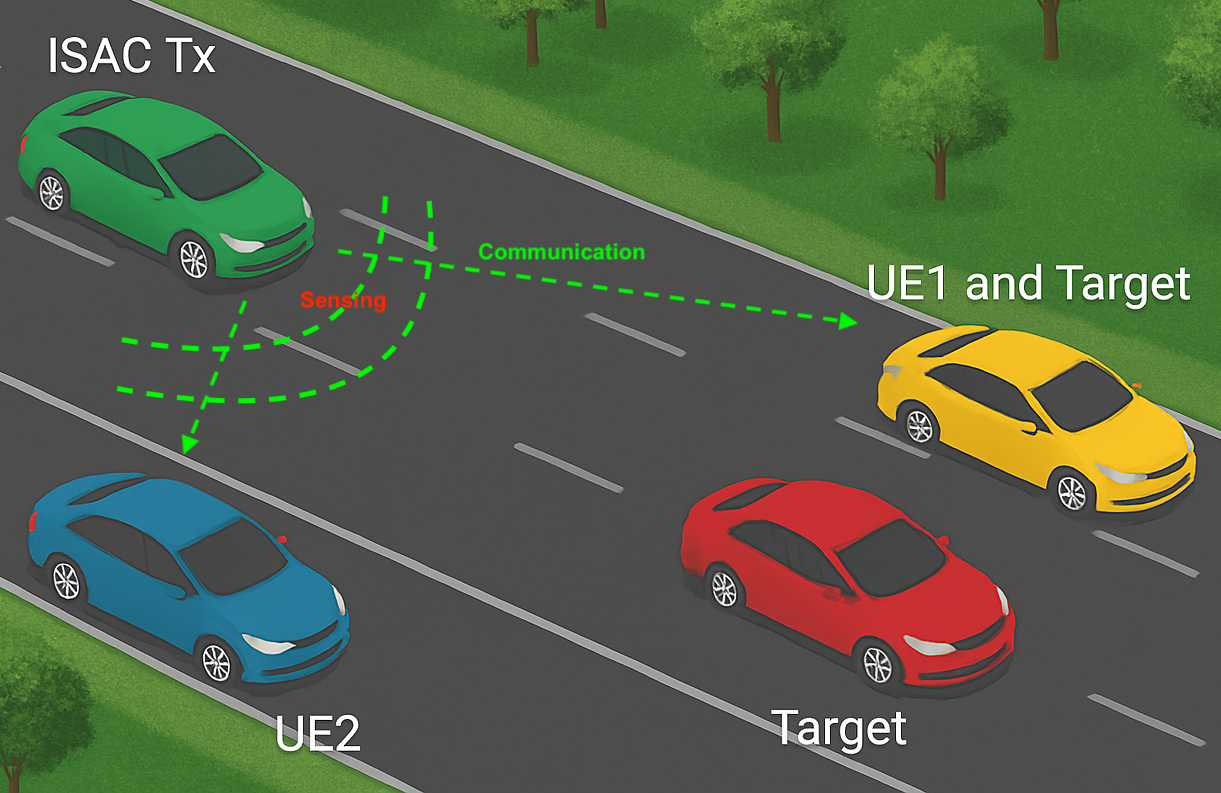}
    \caption{Radar-centric ISAC example: FMCW radar signals are utilized to communicate with other cars while performing sensing.}
    \label{fig:ISAC_case}
\end{figure}

Integrated sensing and communication signaling strategies can be built on the existing communication and sensing waveforms \cite{GirotoJoint2022}. ISAC waveforms based on orthogonal frequency-division multiplexing (OFDM) can be utilized to ensure compatibility with existing communication devices\cite{ Baquero5GRadar2019, TemizDFRC2020, xu2022experimental, TemizOptimizedPrecoders2021}. Such methods are known as communication-centric techniques that aim to use communication signals for radar sensing. OFDM signals are particularly considered for communication-centric ISAC {systems} since they have high flexibility for signal processing and are widely used in modern communication systems {\cite{HsuOFDMISAC2022, FanCross2024}.} {Moreover, other communication waveforms are also considered for ISAC, such as affine frequency division multiplexing (AFDM) waveforms \cite{FanAFDM2026}.} 

On the other hand, using existing radar signals for ISAC is also feasible. Such radar-centric ISAC design modulates data within the radar signals by various modulation techniques, such as index modulation (IM) or phase modulation (PM) to deliver communication data \cite{FanCommunications2019, temiz_2023, Ma_Frac2021, HuangMajorCom2020, Ma_Book2023, AlphanIndex2021, ZhangJCR2021, Li_Joint2023, KafafyV2V2022, HuangISAC2025, WangDFSparse2019, yao2023dual, Xu_DFRC2023, Gu_FH_2022, NiJRC2023}. A radar-centric ISAC system can be implemented within existing radar systems with minimal hardware modification and only with some additional signal processing. Hence, depending on the system's primary aim, enabling both sensing and communication functions within existing systems is feasible by modifying the well-known communication or radar waveforms. The proposed ISAC system in this study is considered under the radar-centric category since it is designed and developed by modifying a radar waveform, namely frequency-modulated continuous wave (FMCW), to transmit data. This study combines PM and IM to enable a high data-rate communication link.

IM has been identified as a highly promising solution for next-generation wireless networks due to its ability to utilize a range of waveform features such as the selection of transmit antennas, subcarriers, and time slots as modulation indices, in contrast to traditional modulation techniques that typically rely on amplitude, frequency, or phase variations to modulate data \cite{BasarIndex2017, MaoNovelIndex2019}. By utilizing several indices simultaneously, large constellation sizes can be achieved, enabling modulation of a larger number of bits into each symbol. IM does not require sophisticated hardware and can be implemented along with other modulation techniques to deliver additional data \cite{Sham_IM2022}.

\input{table1}

Several studies have proposed integrating data transmission within radar systems by modulating sensing waveforms \cite{FanCommunications2019, temiz_2023, Ma_Frac2021, HuangMajorCom2020, Ma_Book2023, AlphanIndex2021, ZhangJCR2021, Li_Joint2023, KafafyV2V2022, WangDFSparse2019, yao2023dual, Xu_DFRC2023, Gu_FH_2022}. For instance, an early study proposed transmitting data by varying the slope of FMCW chirps, where different slopes correspond to distinct data symbols \cite{FanCommunications2019}. Alternatively, Ma et al. introduced a novel approach for data transmission within FMCW systems using index modulation (IM) \cite{Ma_Frac2021}. In another study, Huang et al. proposed a multi-carrier agile joint radar-communication system that leverages antenna and carrier frequency selections of sensing signals to modulate data \cite{HuangMajorCom2020}. Multiple-input-multiple-output (MIMO) antenna systems are also used to design radar-centric ISAC systems, where sensing waveforms are modulated via index, phase, and amplitude modulations to transmit communication data within radar signals \cite{Xu_DFRC2023, Gu_FH_2022}. However, the hardware and signal processing complexity of these methods are high; thus, these studies do not present hardware implementation and measurements. In this study, we propose a low hardware-complexity ISAC system and also present {a proof-of-concept hardware implementation and results of experimental measurements that are performed through a loopback cable.}

In our previous studies \cite{temiz_2023,TemizISACConf2023}, we proposed and developed an IM-based ISAC system that utilizes the fusion of polarization, center frequency, and bandwidth selections of FMCW chirps as indices. A single-channel ISAC transceiver is proposed in \cite{TemizISACConf2023} while a dual-channel ISAC transceiver is proposed in \cite{temiz_2023}. In this study, we have proposed index-modulated phase-coded FMCW (IM-PM-FMCW) signals to increase the system's data rate further while maintaining a high sensing accuracy compared to our previous studies \cite{TemizISACConf2023, temiz_2023}. FMCW waveforms are widely used in automotive radar systems for a range of applications, including advanced driver assistance systems, adaptive cruise control, blind spot detection, and autonomous driving; thus, vehicular systems can be a key application area for the proposed ISAC systems \cite{UysalPCFMCW2020, WangAutomotive2023}. {Hence, this study proposes an FMCW-based ISAC system that can achieve the sensing performance of FMCW radar by turning off data modulation (IM and PM)  when data transmission is not needed.}

A comparison between previous related radar-centric ISAC studies and this study is presented in Table~\ref{table:literature_isac}, where the antenna structures, IM indices, modulation techniques, hardware implementations, hardware and computational complexities, and waveform types are compared. {The main contributions of this study are as follows:}
\begin{itemize}
    \item This study proposes a radar-centric ISAC system that utilizes IM and PM within FMCW chirps to transmit data while maintaining sensing as its primary function. Moreover, it presents a proof-of-concept hardware implementation of the proposed ISAC system and performs experimental measurements to validate its feasibility.
    
    \item It also proposes a transmitter architecture to generate FMCW chirps embedded with IM and PM, along with low-complexity radar and communication receivers. The communication receiver can efficiently demodulate the IM and PM data, while the proposed radar receiver utilizes a novel correction method to mitigate the range and velocity artifacts caused by IM and PM within FMCW chirps. These architectures are fully compatible with current RF and digital front-ends.
    
    \item It also extensively evaluates sensing performance, out-of-band emissions, and communication throughput. This analysis reveals the inherent trade-offs between sensing and communication performances, thereby enabling dynamic waveform optimization based on desired sensing and communication performance requirements.
    
\end{itemize}

\begin{figure}[t]
    \centering
    \includegraphics[width=1\linewidth]{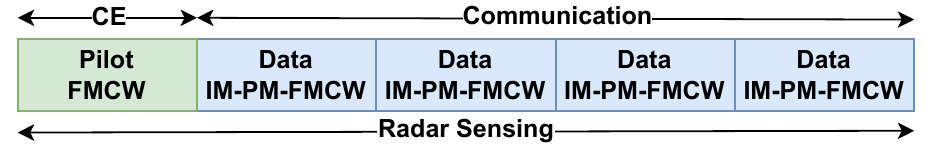}
    \caption{IM-PM-FMCW ISAC frame structure.}
    \label{fig:ce_comm_sense}
\end{figure}

\section{System Model}

In this study, {a vehicular ISAC system is considered as shown in Fig.~\ref{fig:ISAC_case}, where frequency-modulated continuous-wave (FMCW) signals employing index and phase modulations are transmitted to convey communication data to other vehicles while simultaneously performing sensing.} Moreover, vertical (V-pol) and horizontal (H-pol) antenna polarizations are utilized simultaneously to transmit two chirps independently and simultaneously within the same bandwidth, doubling the communication throughput. Index-modulated phase-coded FMCW (IM-PM-FMCW) chirps are digitally generated separately as baseband signals for each polarization. 

The transmission time frame is shown in Fig.~\ref{fig:ce_comm_sense}, where the pilot chirps (FMCW) are transmitted at certain times for channel estimation, as a single chirp transmitted in each polarization is sufficient to estimate the communication channel.  After the transmission of the pilot chirp, a large number of IM-PM-FMCW chirps, carrying the data payload, are transmitted for joint communication and sensing. Radar sensing is not interrupted, as both pilot chirps and data chirps can be used for sensing. {The transmitted frames in V-pol and H-pol are denoted by vectors $\mathbf{x}^V$ and $\mathbf{x}^H$, respectively. Each of the transmitted frames in V-pol and H-pol consists of $I$ chirps, including pilot and data chirps, obtained by concatenating the chirps as,
\begin{align}
    &\mathbf{x}^P = \left[\mathbf{x}_1^P, \dots, \mathbf{x}_i^P, \dots, \mathbf{x}_I^P\right]
\end{align}
where $P \in \{V, H\}$ denotes the polarization index, i.e., vectors $\mathbf{x}_i^V \in \mathbb{C}^{1 \times T_c}$ and $\mathbf{x}_i^H \in \mathbb{C}^{1 \times T_c}$ denote the $i$-th chirps transmitted in V-pol and H-pol, respectively, and they are given by
\begin{align}
    &\mathbf{x}_i^P = \left[x^P_1(1), \dots, x^P_i(t), \dots, x^P_i(T_c)  \right]
\end{align} where $x^P_i(t)$ denotes the complex-valued samples at time $t$ with $0\leq t < T_c$.}

\subsection{Index Modulation}
{Let $x_i(t)$ denote the $t$-th sample of the $i$-th chirp in one polarization}, which is given by,
\begin{equation}
x^P_i(t)=\cos\left(\theta^P_i(t)+\phi_{re}^{P,l}(t)\right)+j\sin\left({\theta^P_i(t)+\phi_{im}^{P,l}(t)}\right), \label{fmcw_cis}
\end{equation}
where $\theta^P_i(t)$ denotes the time-dependent phase of the chirp at time $t$, and $\phi_{re}^{P,l}(t)$ and $\phi_{im}^{P,l}(t)$ denote the phases of the real and imaginary part of the baseband signal for each $l$th segment of the chirp consisting of $L$ segments, such that $l = 1, 2, \dots, L$ and transmitted in V-pol or H-pol, i.e., $P\in\{H,V\}$. 

In $x^P_i(t)$ given by (\ref{fmcw_cis}), the time-dependent phase $\theta_i(t)$ can be expressed in terms of the chirp bandwidth and the center frequency. Let $b^P_i$ and $f^P_i$ denote the bandwidth and center frequency of the $i$-th chirp, corresponding to the IM data in one polarization. In this case, the start and end frequency of the $i$-th chirp is given by $f_{i,0}^P = f^P_i-b^P_i/2$ and $f_{i,1}^P = f^P_i+b^P_i/2$. Hence, the time-dependent phase of the chirp, $\theta_i(t)$, is given  by,
\begin{equation}
    \theta^P_i(t) = \pi\left[\frac{f_{i,1}^P-f_{i,0}^P}{2}t^2+2f_{i,0}^Pt\right], \quad 0\leq t < T_c,\label{eq_angle}
\end{equation} by replacing $f_i$ and $b_i$ in (\ref{eq_angle}), the following equation is obtained,
\begin{equation}
    \theta_i(t) = \pi\left[\frac{b_i}{2}t^2+2(f_i-\frac{b_i}{2})t\right], \quad 0\leq t < T_c,\label{eq_angle_2}
\end{equation}
where $T_c$ denotes the chirp duration.  Thus, the chirp center frequency $f_i$ and bandwidth $b_i$ are indices of the IM. Considering that a dual-polarized antenna is employed, they can have independent values in each polarization; accordingly, $[f_i^V, b_i^V]$ and $[f_i^H, b_i^H]$ can be written for V-pol and H-pol transmissions.

\subsection{Phase Modulation}\label{sec:phase_mod}

The single chirp given by (\ref{fmcw_cis}) is divided into $L$ segments, and the duration of each of them is $\frac{T_c}{L}$, corresponding to $L$ distinct phases within a chirp. The phase of each segment is denoted by $\phi_{re}^l(t)$ and $\phi_{im}^l(t)$. The phase of the real and imaginary parts of the signal is kept the same for each $l$th segment, hence $\phi_{re}^l(t) = \phi_{im}^l(t) = \phi_i^l(t)  $. Moreover, independent phase codes\footnote{The sequence of phase modulations within a chirp, i.e., phases of $L$ segments of a chirp, is defined as phase codes in line with the previous radar studies \cite{UtkuPCFMCW2023}.} can be transmitted in each polarization, hence, $\phi_i^{V,l}(t)$ and $\phi_i^{H,l}(t)$ are defined for V-pol and H-pol transmissions. Hence, the transmitted phases during the $i$-th chirp duration in V-pol and H-pol are respectively given by
\begin{equation}
    \Phi_i^V = \left[\phi_i^{V,1}, \phi_i^{V,2},\dots, \phi_i^{V,L}\right],
\end{equation}
\begin{equation}
    \Phi_i^H = \left[\phi_i^{H,1}, \phi_i^{H,2},\dots, \phi_i^{H,L}\right],
\end{equation}
{where the phase symbols belong to an $M$-PSK alphabet, i.e.,
\begin{equation}
\phi_i^{V,l},\,\phi_i^{H,l}\in\left\{0,\frac{2\pi}{M},\ldots,\frac{2\pi(M-1)}{M}\right\},
\qquad l=1,2,\dots,L.
\end{equation}
Using higher-order phase modulation results in higher data rates.} Assuming that the order of the PM in each phase segment is denoted by $M$, the total data transmitted via PM in each chirp duration is $2L \log_2(M)$ bits by utilizing a dual-polarized antenna.

{
To mitigate out-of-band (OOB) spectral leakage caused by abrupt phase transitions at the chip boundaries, a continuous-phase smoothing filter is applied. Let $T_s = T_c/L$ denote the duration of a phase segment, and the normalized truncated Gaussian impulse response with span $\beta T_s$ is defined as
\begin{equation}
h_t(t)=
\begin{cases}
C \exp\!\left(-\dfrac{t^2}{2\sigma^2}\right), & |t|\le \dfrac{\beta T_s}{2},\\[6pt]
0, & \text{otherwise},
\end{cases}
\end{equation}
where $\beta\in(0,1]$ controls the truncation span, $\sigma$ dictates the effective (3-dB) bandwidth, and
$C$ is chosen such that $\int_{-\beta T_s/2}^{\beta T_s/2} h_t(t)\,dt=1$. To reduce OOB leakage while limiting
phase distortion, we set $\beta=0.2$, so the phase transitions glide over $0.2T_s$, instead of sudden phase changes.

The continuous-time phases over
$1\le t < T_c$ within a chirp transmitted in V-pol or H-pol are given by
\begin{equation}
\phi_i^P(t)=\sum_{l=1}^{L}\phi_i^{P,l}\Big[u\!\big(t-(l-1)T_s\big)-u\!\big(t-lT_s\big)\Big],
\end{equation}
where $u(\cdot)$ denotes the unit-step function. The continuous-time smoothed phases for V-pol and H-pol transmissions are obtained by
\begin{align}
\tilde{\phi}_i^P(t) &= \arg\!\left(\big(e^{j\phi_i^P(t)}\big) * h_t(t)\right),
\label{eq:smoothed_phases}
\end{align}
where $\arg(.)$ and $*$ denote the phase of a complex number and linear convolution, respectively.

By utilizing the smoothed phases in \eqref{eq:smoothed_phases} and Euler's formula, baseband IM--PM--FMCW chirps for V-pol or H-pol transmissions are denoted by
\begin{equation}
x_i^P(t) = \exp\!\left(-j \pi\left[\frac{b_i^P}{2}t^2+2\!\left(f_i^P-\frac{b_i^P}{2}\right)t\right]
+ j\tilde{\phi}_i^{P}(t)\right),
\end{equation} where $0\leq t < T_c$ and $l=1, 2, \dots, L$ for each chirp, $x_i^P\in\{x_i^V,x_i^V\}$, $b_i^P\in\{b_i^V,b_i^H\}$, $f_i^P\in\{f_i^V,f_i^H\}$. Accordingly, the transmitted data codewords for the $i$th chirp duration by utilizing both polarizations are given by
\begin{equation}
    \Psi_i = \left[f_i^V, b_i^V, f_i^H, b_i^H, \Phi_i^V, \Phi_i^H \right],
\end{equation}
where each parameter is drawn from a modulation codebook.}

\section{Joint ISAC Transmitter and Receiver Architecture}\label{Sec:TxRXArc}
\begin{figure*}
    \centering
    \includegraphics[width=0.75\linewidth]{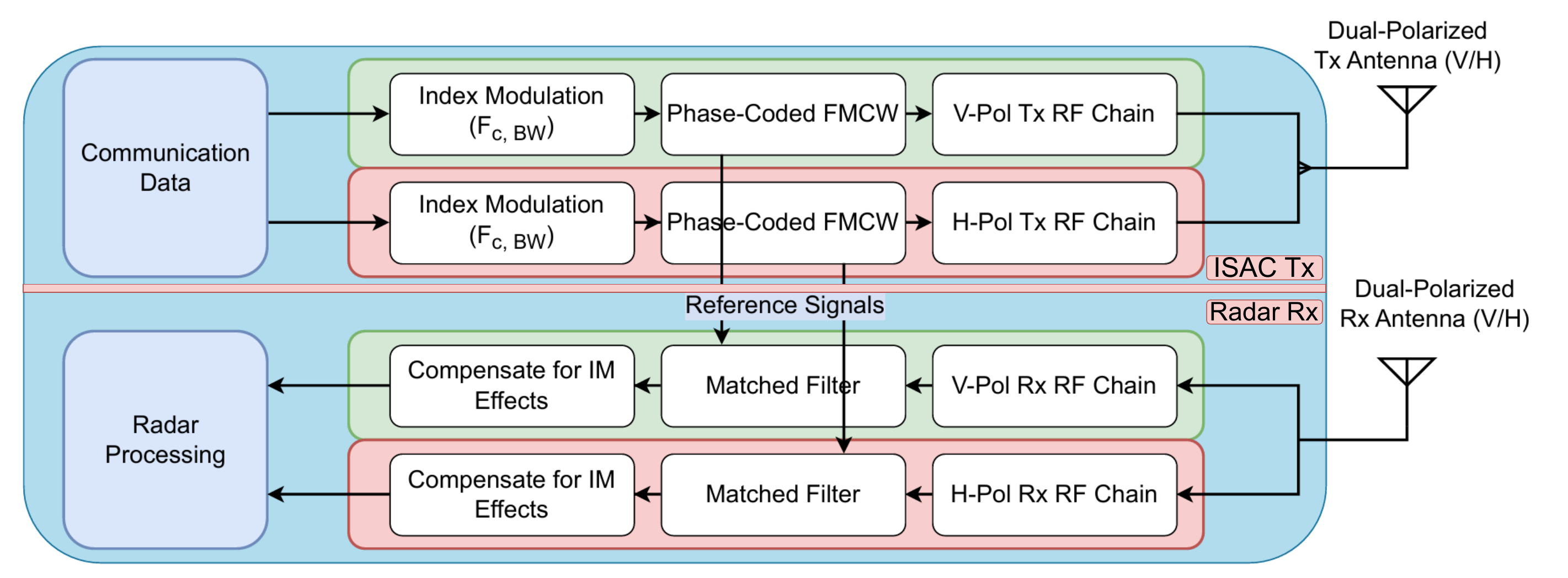}
    \caption{Joint ISAC transmitter and radar receiver architecture for IM-PM-FMCW waveform.}
    \label{fig:tx_model}
\end{figure*}

The proposed joint ISAC transmitter and radar receiver architecture for the IM-PM-FMCW ISAC system is shown in Fig.~\ref{fig:tx_model}, which modulates the FMCW signals via IM and PM with the communication data that is to be delivered to the communication users that know the modulation codebook. In addition, its radar receiver receives and processes the reflected signals from the targets in full-duplex mode for sensing.

\begin{figure*}[h]
    \centering \includegraphics[width=0.75\linewidth]{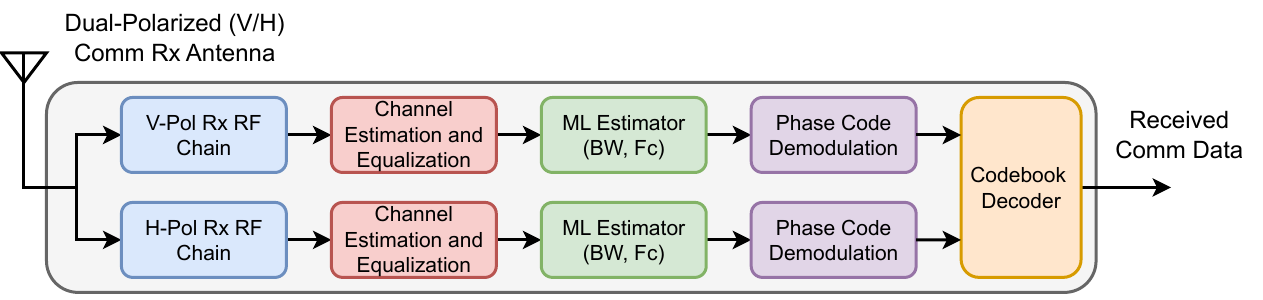}
    \caption{The proposed dual-polarized communication receiver architecture to receive and demodulate the signals transmitted by the proposed ISAC transmitter.}
    \label{fig:rx_com_model}
\end{figure*}

In the ISAC transmitter shown in Fig.~\ref{fig:tx_model}, the bandwidth and center frequency of the chirps are first decided by IM data. Then, the phase-coded FMCW chirps that carry data with the PM are generated with these bandwidths and center frequencies. The digitally generated index-modulated and phase-coded FMCW chirps are then processed through the V-pol and H-pol RF chains and are transmitted by V-pol and H-pol transmit antennas. 

The indices of IM are determined as the bandwidth of FMCW chirps and the center frequency of chirps. The frequency of the carrier signal is denoted by $F_c$, and changing the center frequency of the chirp from $0$ to $f_i$ will shift the chirp to $F_c+f^V_i$ or $F_c+f^H_i$ for V-pol and H-pol in the RF domain, providing a frequency-hopping radar signal.  Thus, it is a frequency-hopping radar that also inherently provides a low probability of intercept. In practical scenarios, the interference between FMCW radars can adversely affect the received signals, leading to a loss in target detection and parameter estimation accuracy. To overcome this issue, the phase-coded FMCW (PC-FMCW) radar employs phase codes to distinguish between the signals \cite{UtkuPCFMCW2023, UysalPCFMCW2020}. Accordingly, the proposed method is also expected to reduce the interference between radars operating in the vicinity of each other.

\subsection{Radar Receiver Processing}\label{Section:Radar_processing}

In the monostatic radar receiver, the second chain shown in Fig.~\ref{fig:tx_model}, the received signals, which are reflected from the targets, are down-converted by the RF chains of V-pol and H-pol chains and then digitized by analog-to-digital converters (ADCs). Deramping and decimation of the signals are digitally performed, and then the FFT of the resulting signals is taken to obtain the target range information.

Assuming a single target is present, the received baseband signal {sample at time $t$} at the radar receiver in each polarization is given by
\begin{equation}
    r_i(t) =  x_i(t-\tau) e^{j2\pi (F_c+f_i)(t-\tau)} e^{j2\pi f_D t} {+ n(t)},
\end{equation}
where $\tau={2R_T}{/c}$ is the delay introduced by the target range $R_T$, and the $f_D=2{V_T}/{\lambda}$, where $\lambda$ denotes the wavelength of the signal, is the Doppler shift caused by the target velocity $v_T$. Moreover, $n(t)$ denotes the complex-valued additive white Gaussian noise (AWGN) sample with zero mean and noise variance $\sigma_n^2$ as  $\mathcal{CN}\left(0,\sigma_n^2\right)$.

At the receiver, the mixing of signals is performed as $s_i(t) = r_i(t)x_i^*(t)$. {Taking the fast Fourier transform (FFT) of $s_i(t)$ produces the delay-power map, corresponding to the range profile of the target for standard FMCW chirps, }{ since the bandwidth and center frequency remain constant across all chirps of a standard FMCW radar. However, in the proposed IM-PM-FMCW scheme, the $i$-th chirp is modulated with a specific bandwidth $b_i$ and center frequency $f_i$ to embed communication data. Thus, to mitigate range variations and slow-time phase corruption during target estimation, we have proposed a range and Doppler compensation method as explained below.}

Assuming that the range domain is discretized into a uniformly spaced estimation vector as
\begin{equation}
    \mathbf{r} = [\,r_0,\; r_1,\; \dots,\; r_{N_r-1}\,]^T,
\end{equation}
where $N_r$ is the number of range bins. For each hypothesized range bin $r_m \in \mathbf{r}$, the corresponding two-way propagation delay is $\tau(r_m) = {2r_m}/{c}$,
where $c$ denotes the speed of light.

\subsubsection{Bandwidth Compensation (Range Alignment)}
Because the bandwidth $b_i$ varies from chirp to chirp, the chirp slope, $S_i = {b_i}/{T_c}$,
is also chirp-dependent. As a result, a target at a fixed range produces a different beat frequency in each chirp. For the $i$-th chirp, the beat frequency corresponding to range bin $r_m$ is given by
$f_{\mathrm{beat},i}(r_m) =  {2S_i r_m}/{c}.$ Let $y_i(n)$ denote the sampled fast-time beat signal of the $i$-th chirp. 

{To correct the range-bin misalignment, a chirp-dependent discrete-time Fourier transform (DTFT) evaluation is used in place of the conventional FFT-based range mapping, such that each target is sampled at the beat frequency corresponding to its range in the given chirp. This ensures that the target energy is aligned to the same range bin across all chirps. Accordingly, the range response at bin $r_m$ is obtained by evaluating the fast-time beat signal at the corresponding chirp-dependent beat frequency, as
\begin{equation}
    R_i(r_m)
    = \sum_{n=0}^{N_s-1} y_i(n)\,
    e^{-j2\pi f_{\mathrm{beat},i}(r_m)nT_s},
\end{equation}
where $T_s$ and $N_s$ denote the fast-time sampling interval and the number of samples per chirp, respectively. Repeating this operation for all $r_m$ forms the range profile of the $i$-th chirp,
\begin{equation}
    \mathbf{r}_i =
    [\,R_i(r_0),\; R_i(r_1),\; \dots,\; R_i(r_{N_r-1})\,]^T.
\end{equation}
This step aligns the target energy to a common range bin across all $N_c$ chirps and mitigates range migration caused by bandwidth hopping.}

{\subsubsection{Phase Error Correction (Doppler Alignment)}
After range alignment, the index modulation still introduces deterministic slow-time phase errors that blur the Doppler response. The following two terms must be compensated,
\begin{itemize}
    \item {Linear Phase Error:} The chirp-to-chirp center frequency variations $\Delta f_i = f_i - f_{\text{ref}}$, where $f_{\text{ref}}$ is the reference center frequency, introduce a phase error proportional to the target delay.
    \item {Quadratic Phase Error:} The chirp-to-chirp bandwidth variations introduce a bandwidth-dependent quadratic phase error.
\end{itemize}
Assume that the chirp-slope deviation for the $i$-th chirp is defined by $\Delta S_i = \left({b_i - b_{ref}}\right)/{T_c}$,  where $b_{ref}$ denotes the reference bandwidth. Then, the total deterministic phase error at range bin $r_m$ for the $i$-th chirp is given by
\begin{equation}
    \phi_{\mathrm{err},i}(r_m)
    = -2\pi \Delta f_i \tau(r_m)
    + \pi \Delta S_i \tau^2(r_m), 
\end{equation} where the first term, $-2\pi \Delta f_i \tau(r_m)$, is caused by the center-frequency offset and is linear in the target delay, while the second term, $\pi \Delta S_i \tau^2(r_m)$, is caused by the chirp-slope deviation and is quadratic in the delay. 
Thus, to compensate for this error, the corresponding phase-correction factor is given by $p_i(r_m) = e^{-j\phi_{\mathrm{err},i}(r_m)}$.
Since the radar receiver knows both chirps' centre frequency and bandwidth, it can compensate for this error as, 
\begin{equation}
    \tilde{R}_i(r_m) = R_i(r_m)\, p_i(r_m)
    = R_i(r_m)\, e^{-j\phi_{\mathrm{err},i}(r_m)}.
\end{equation}}

{\subsubsection{Range-Doppler Processing}
After phase correction, the slow-time sequence at each fixed range bin follows the true Doppler phase progression, decoupled from the communication-induced phase distortion. For a given range bin $r_m$, the compensated responses from all chirps are written as
\begin{equation}
    \tilde{R}_0(r_m),\; \tilde{R}_1(r_m),\; \dots,\; \tilde{R}_{N_c-1}(r_m).
\end{equation} Then, a standard FFT is applied across the chirp index to obtain the Doppler spectrum at each range bin as,
\begin{equation}
    D(r_m,k)=\mathrm{FFT}\{\tilde{R}_i(r_m)\},
\end{equation} where $k$ denotes the Doppler-bin index of the slow-time FFT. Repeating this operation for all range bins yields the final focused range-Doppler map.}

{The proposed method is dominated by the DTFT-based range-alignment stage, with overall complexity $\mathcal{O}(N_c N_r N_s)$. The phase-correction and Doppler-FFT stages add $\mathcal{O}(N_c N_r)$ and $\mathcal{O}(N_r N_c \log N_c)$, respectively. By comparison, standard FMCW processing has complexity $\mathcal{O}(N_c N_s \log N_s + N_c N_s \log N_c)$, since it uses an $N_s$-point FFT in fast time and an $N_c$-point FFT in slow time. Hence, the proposed method is computationally more demanding due to the explicit DTFT evaluation over all range bins. However, the additional complexity remains within limits that can be handled by the hardware. Accordingly, the proposed radar processing can employ the conventional FMCW receiver chain with low-resolution ADCs, but it requires an increased computational complexity due to the additional range- and velocity-compensation steps.}

{\section{Communication Channel Model and Receiver Architecture} This section details the vehicular communication channel model under consideration, alongside the receiver architecture, which encompasses both channel estimation and the demodulation of IM and PM data.}

\subsection{Communication Channel Model and Channel Estimation}
{Since a vehicular scenario is considered, as illustrated in Fig.~\ref{fig:ISAC_case}, a Rician fading channel model is adopted to capture the presence of a dominant line-of-sight (LOS) component together with multipath scattering. Moreover, the Doppler shift and cross-polarization interference are also considered in this channel model since it employs dual-polarized antennas. In our previous field trials and measurements with dual-polarized antennas, we have seen around - 14 dB polarisation leakage between V-pol and H-pol channels in the measured data at the receiver \cite{temiz_2023}. Thus, only the LOS path of the cross-polarization interference channel is included in the model, for the sake of simplicity.}

The wideband signals received in the vertical (V-pol)  polarization channel at the communication receiver during the $i$-th chirp are given by,
{\begin{align}
r^V_i(t) &= \underbrace{\sqrt{\frac{K_R}{K_R+1}}\, h^V_{\text{LOS}}\, x^V_i(t-\tau_0)\, e^{j 2\pi \nu_0 t}}_{\text{LOS Component}} \notag \\
&+\underbrace{\sqrt{\frac{1}{K_R+1}} \sum_{l=1}^{L} h^V_l\, x^V_i(t-\tau_l)\, e^{j 2\pi \nu_l t}}_{\text{NLOS Multipath Component}} \nonumber\\
&+ \underbrace{h^{HV}_0\, x^H_i(t-\tau_0)\, e^{j 2\pi \nu_0 t}}_{\text{LOS Cross-pol Component}} + n(t), \label{channelV}
\end{align}
where $K_R$ denotes the Rician $K$-factor, defined as the ratio of the average power of the LOS component to that of the scattered non-line-of-sight (NLOS) components. 
The index $l=0$ corresponds to the LOS path with delay $\tau_0$, and Doppler shift $\nu_0$, while $l=1,\dots,L$ represent the $L$ NLOS multipath components with delays $\tau_l$ and Doppler shifts $\nu_l$. 
The coefficient $h^V_{\text{LOS}}$ denotes the LOS channel gains for the V-pol channel, whereas $h^V_l \sim \mathcal{CN}(0,1)$ models the random complex gains of the NLOS components.  The cross-polarization coupling coefficient $h^{HV}_l$ accounts for polarization leakage from H-pol to V-pol. Cross-polarization interference between two polarization channels occurs due to RF signal leakage between the polarizations and reflections in the propagation environment.  Moreover, $n(t)$ denotes the AWGN with zero mean and noise variance $\sigma_n^2$ as  $\mathcal{CN}\left(0,\sigma_n^2\right)$. The V-pol channel is also modeled in the same way by considering the cross-polarization leakage from V-pol to H-pol.}

{%
Channel estimation and equalization are performed at the communication receiver, as shown in Fig.~\ref{fig:rx_com_model}, before demodulating IM and PM, since the channel introduces random phase shifts, amplitude variations, and inter-symbol interference. Standard FMCW pilot chirps are transmitted for channel estimation, and the subsequent IM-PM-FMCW chirps are equalized using the obtained channel state information (CSI). The received frequency-domain pilot signals in the V-pol and H-pol are given by
\begin{align}
\mathbf{y}_p^V &= \mathbf{h}^V\odot\mathbf{u}_p^V+\mathbf{h}^{HV}\odot\mathbf{u}_p^H+\mathbf{n}^V,\notag \\
\mathbf{y}_p^H &= \mathbf{h}^H\odot\mathbf{u}_p^H+\mathbf{h}^{VH}\odot\mathbf{u}_p^V+\mathbf{n}^H,
\end{align}
respectively, where $\mathbf{u}_p^V=\mathcal{FFT}(\mathbf{x}_p^V)$ and $\mathbf{u}_p^H=\mathcal{FFT}(\mathbf{x}_p^H)$ are the transmitted pilot FMCW chirps represented in the frequency domain. The operators $\odot$ and $\oslash$ denote element-wise multiplication and division, respectively. The bandwidth of pilot chirps occupies the entire allocated bandwidth to estimate the entire channel response of the allocated bandwidth for IM-PM-FMCW chirps.}

{%
To suppress the impact of noise and cross-polarization interference with low complexity, we employ a frequency-domain linear minimum mean square error (LMMSE) channel estimator applied element-wise over the FFT bins. Let $\mathbf{I}_1\in\mathbb{R}^{N_s\times 1}$ denote an all-ones vector, where $N_s$ denotes the number of samples within a chirp. Assuming identical noise and polarization statistics in both polarizations, with thermal noise variance $\sigma_n^2$ and cross-polarization interference variance $\sigma_i^2$, the channels in V-pol and H-pol are estimated as
\begin{align}
\mathbf{h}_e^V &=
\left(\mathbf{y}_p^V\odot(\mathbf{u}_p^V)^*\right)\oslash
\left(\mathbf{u}_p^V\odot(\mathbf{u}_p^V)^*+(\sigma_n^2+\sigma_i^2)\mathbf{I}_1\right), \notag\\
\mathbf{h}_e^H &=
\left(\mathbf{y}_p^H\odot(\mathbf{u}_p^H)^*\right)\oslash
\left(\mathbf{u}_p^H\odot(\mathbf{u}_p^H)^*+(\sigma_n^2+\sigma_i^2)\mathbf{I}_1\right),
\label{eq:lmmse_ce}
\end{align}
where $(\cdot)^*$ denotes complex conjugation.}

{%
This estimated CSI is utilized to equalize subsequent IM-PM-FMCW chirps until a new pilot FMCW chirp is received, and the channel estimate is updated. Let $\mathbf{y}_i^V$ and $\mathbf{y}_i^H$ denote the received frequency-domain signals for the $i$th IM-PM-FMCW data chirp. The data chirps are equalized using a minimum mean square error (MMSE) equalizer, which is applied element-wise as
\begin{align}
\hat{\mathbf{u}}_i^V &=
\left((\mathbf{h}_e^V)^*\odot\mathbf{y}_i^V\right)\oslash
\left(\mathbf{h}_e^V\odot(\mathbf{h}_e^V)^*+(\sigma_n^2+\sigma_i^2)\mathbf{I}_1\right), \notag\\
\hat{\mathbf{u}}_i^H &=
\left((\mathbf{h}_e^H)^*\odot\mathbf{y}_i^H\right)\oslash
\left(\mathbf{h}_e^H\odot(\mathbf{h}_e^H)^*+(\sigma_n^2+\sigma_i^2)\mathbf{I}_1\right).
\label{eq:mmse_eq}
\end{align}
Finally, the time-domain equalized chirps are obtained as $\hat{\mathbf{x}}_i^V=\mathcal{IFFT}(\hat{\mathbf{u}}_i^V)$ and $\hat{\mathbf{x}}_i^H=\mathcal{IFFT}(\hat{\mathbf{u}}_i^H)$.}
\subsection{Communication Receiver}

Fig.~\ref{fig:rx_com_model} illustrates the block diagram of the proposed communication receiver architecture, where V-pol and H-pol antennas receive the IM-PM-FMCW chirps. Then, the received RF signals are processed by receiver RF chains (which can be entirely digital, as in the UCL`s ARESTOR platform discussed in the following, or conventional RF chains) and converted to baseband signals. After that, digitized baseband signals are processed for channel estimation and equalization. 

Pilot chirps (FMCW chirps) in both polarizations occupying the entire available frequency are transmitted for time synchronization and channel estimation. FMCW chirps are also used for sensing, as shown in Fig.\ref{fig:ce_comm_sense}. Hence, the proposed architecture can perform radar sensing at all times by employing pilot chirps (FMCW) and data-carrying IM-PM-FMCW chirps.

\subsection{Demodulation of Index Modulation}\label{demod_im_section}
\begin{figure}
    \centering
    \includegraphics[width=0.9\linewidth]{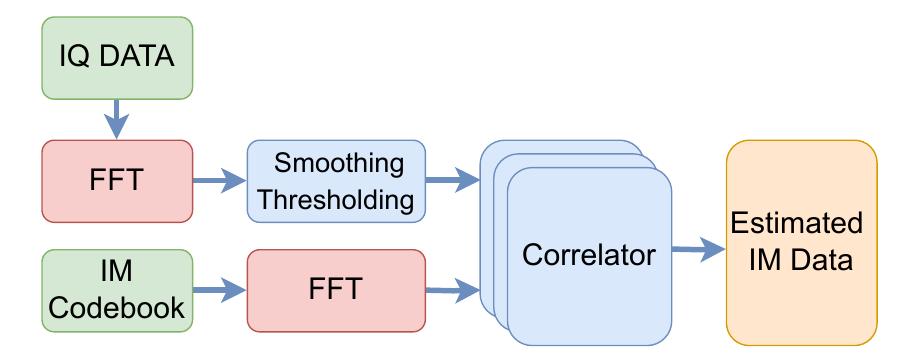}
    \caption{Signal processing for the demodulation of IM.}
    \label{fig:IM_demod2}
\end{figure}
After channel estimation and equalization, the communication receiver estimates bandwidth ($b_i$) and center frequency indices ($f_i$) for each chirp received. Two maximum likelihood (ML) estimators concurrently perform in the frequency domain, as shown in Fig. \ref{fig:rx_com_model}; each estimator is connected to one of the dual-polarised antenna channels (V-pol and H-pol antennas). The receiver knows the IM codebook consisting of chirps with all possible center frequencies and bandwidths as
\begin{equation}
\mathcal{F}_{IM} = \{\mathbf{u}_{(1,1)}, \dots \mathbf{u}_{(b,f)}, \dots, \mathbf{u}_{(N_{b},N_{f})}\}, 
\end{equation}
where $\mathcal{F}_{IM}$ denotes the IM codebook, $N_{b}$ and $N_{f}$ denote the number of bandwidth and centre frequency options. Using the IM codebook, two ML estimators can concurrently operate to detect the symbols transmitted in V-pol and H-pol as,
\begin{argmaxi}
    {(\hat{b}^V_i,\hat{f}^V_i)}{|\mathbf{u}^V_{i} \mathbf{u}^*_{(b,f)} |^2}
    {}{}
    ,\label{MLestimator2a2}
\end{argmaxi}
\begin{argmaxi}
    {(\hat{b}^H_i,\hat{f}^H_i)}{|\mathbf{u}^H_{i} \mathbf{u}^*_{(b,f)} |^2}
    {}{}
    ,\label{MLestimator2b2}
\end{argmaxi}
where $*$ denotes the complex conjugate, and the ML estimators return the estimated indices $(\hat{b}^V_i,\hat{f}^V_i)$ and $(\hat{b}^H_i,\hat{f}^H_i)$ for the $i$-th chirps received in V-pol and H-pol.  Consequently, the center frequency and bandwidth of the $i$-th chirp are estimated. 

The implementation of the ML estimators used for IM demodulation is illustrated by Fig.~\ref{fig:IM_demod2}, which is performed in the frequency domain. To improve the accuracy of IM demodulation, frequency domain data is normalized, and a low-pass filter is applied to the frequency domain data to smooth the frequency response. The computational complexity of each ML estimator is related to the size of the IM codebook, i.e., $N_{b}N_{f}$. 
After estimating the indices of the chirps using the dual ML IM estimators mentioned above, these indices are remapped to the communication data, corresponding to received IM bits.

\subsection{Demodulation of Phase Modulation}\label{Sec:Demod_PC}

Before demodulating the PM, it is necessary to correctly demodulate the IM so that the communication receiver can know the center frequency and bandwidth of the chirps. After taking IFFT of $\hat{\mathbf{u}}_i^V$ and $\hat{\mathbf{u}}_i^H$, channel-equalised time domain signal vectors $\hat{\mathbf{x}}_i^V$ and $\hat{\mathbf{x}}_i^H$ are obtained. After that, reference signals with the same center frequency and bandwidth but different phases are generated and multiplied by the received chirp. Then, the phase codes of the chirp are estimated by deciding the maximum matching for each $l$th section of the chirp, where 4 phase options are considered for each phase transition, i.e., ${\phi_m} \in [0, \pi/2, \pi, 3\pi/2]$.  The vectors of reference signals with different phases for V-pol (the same operation is also concurrently performed for the H-pol) are generated by
\begin{align}
           \mathbf{x}_{i,m}^V(t) =\cos({\hat{\theta}^V_i(t)+\phi_m})+j\sin({\hat{\theta}^V_i(t)+\phi_m}), 
\end{align}
\begin{align}
           \mathbf{x}_{i,m}^H(t) =\cos({\hat{\theta}^H_i(t)+\phi_m})+j\sin({\hat{\theta}^H_i(t)+\phi_m}), 
\end{align}
where the frequency sweep, $\hat{\theta}_i^V(t) $, is defined by the following equations for V-pol and H-pol, 
\begin{equation}
    \hat{\theta}_i^V(t) = \pi\left[\frac{\hat{b}_i^V}{2}t^2+2(\hat{f}_i^V-\frac{\hat{b}_i^V}{2})t\right], \label{ref_angle}
\end{equation}
\begin{equation}
    \hat{\theta}_i^H(t) = \pi\left[\frac{\hat{b}_i^H}{2}t^2+2(\hat{f}_i^H-\frac{\hat{b}_i^H}{2})t\right], \label{ref_angle}
\end{equation}
based on the estimated center frequencies $\hat{f}^V_i$, $\hat{f}^H_i$, and bandwidths $\hat{b}^V_i$, $\hat{b}^H_i$ of the $i$-th chirp in the V-pol and H-pol using estimators given by (\ref{MLestimator2a2}) and (\ref{MLestimator2b2}). Then, the received chirp is multiplied by the complex conjugate of the reference chirps. The receiver integrates the multiplication results for each phase segment, $l=1, 2, \dots, L$, of the chirp and estimates the phases within a chirp for V-pol and H-pol as, 
\begin{argmaxi}
    {{m_l}^V}{|\mathbf{x}^V_{i,l} \mathbf{x}^{V*}_{i,m,l}|^2}
    {}{}
    ,\label{MLestimator2a2Phase}
\end{argmaxi}
\begin{argmaxi}
    {m_l^H}{|\mathbf{x}^H_{i,l} \mathbf{x}^{H*}_{i,m,l} |^2}
    {}{}
    ,\label{MLestimator2b2Phase}
\end{argmaxi}
where vectors $\mathbf{x}^H_{i,l}$ and $\mathbf{x}^{H*}_{i,m,l}$ denote the $l$th section of vectors $\mathbf{x}^H_{i}$ and $\mathbf{x}^{H*}_{i,m}$ for $l=1,2,\dots, L$. Estimated phase vectors are then obtained as $\mathbf{m}^V = [m_1^V, m_2^V,\dots, m_L^V]$ and $\mathbf{m}^H = [m_1^H, m_2^H,\dots, m_L^H]$ for V-pol and H-pol channels. After completing demodulation of IM and PM, the final data is obtained by getting the corresponding data from the codebook for the estimated IM and PM data ($\hat{b}^V_i,\hat{f}^V_i, \mathbf{m}^V$) for V-pol and ($\hat{b}^H_i,\hat{f}^H_i, \mathbf{m}^H$) for H-pol.

{For a single chirp, the demodulation complexity of the proposed receiver is given by $\mathcal{O}(N_{b} N_{f} N_s + MN_s)$, where $N_b$ and $N_f$ are the numbers of bandwidth and center-frequency states, respectively.}

\begin{figure*}
    \centering
    \includegraphics[width=0.7\linewidth]{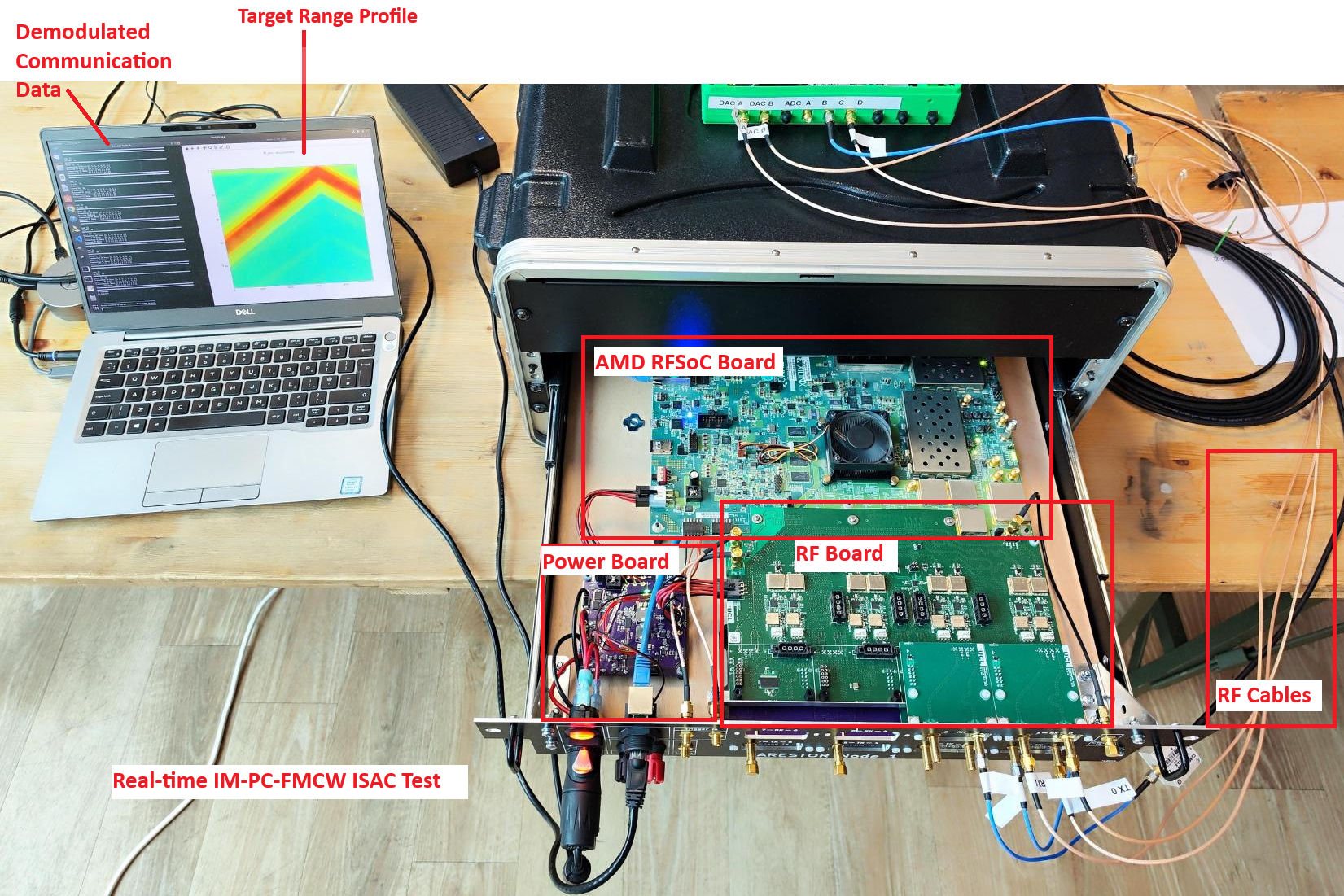}
    \caption{Experimental measurements and real-time signal processing, where UCL's ARESTOR is used as ISAC transmitter, radar receiver, and communication receiver via a loopback cable since it has multiple RF channels. The real-time signal processing is performed in ARESTOR's FPGA and ARM CPU, and the results are shown on a computer screen.}
    \label{fig:exp_real_time}
\end{figure*}

\section{Throughput Analysis}
The proposed ISAC system utilizes a single dual-polarized antenna; hence, two independent chirps can be simultaneously transmitted by V-pol and H-pol antennas. This doubles the data rate that can be transmitted during each chirp duration. Consequently, the maximum data rate that can be transmitted is given by,
\begin{equation}
    R_{max} = 2 F_r \left[N_{IM} + N_{P}\right] \quad bits/s, 
\end{equation}
where $F_r$ denotes the chirp repetition frequency, which is the inverse of chirp duration, as $F_r=N_{IM}/{T_c}$ denotes the number of bits transmitted per chirp via IM as
\begin{equation}
    N_{IM} = \lfloor \log_2\left(N_{b}N_{f} \right)\rfloor,\label{im_bits_size}
\end{equation}
where $\lfloor.\rfloor$ defines the floor operation, $N_{b}$ denotes the number of bandwidth options, and $N_{f}$ denotes the number of center frequency options, from which the bandwidth and center frequency combinations of the chirps can be drawn. Accordingly, the size of the IM constellation can be given by $(N_{b}N_{f})$. Moreover, $N_p$ denotes the number of bits that are transmitted by phase modulation in each chirp in each polarisation, and it is given by,
\begin{equation}
    N_p = L \log_2(M),
\end{equation}
where $L$ denotes the number of phase transitions along a chirp and $M$ denotes the modulation order of the PM. Accordingly, the maximum data rate is given by,
\begin{equation}
    R_{max} = \frac{2\lfloor \log_2\left(N_{b}N_{f} \right)\rfloor +2 L \log_2(M)}{T_c}.
\end{equation}
Assuming that symmetrical demodulation performances are achieved in both polarisations, i.e., the block error rates (BLER) of V-pol and H-pol transmissions are the same, instantaneous throughput is given by, 
\begin{equation}
    R_i = \frac{2(1-\gamma_I)\lfloor \log_2\left(N_{b}N_{f} \right)\rfloor +2(1-\gamma_P) L \log_2(M)}{T_c}, \label{ins_rate}
\end{equation}
where $\gamma_I$ and $\gamma_P$ denote the block error rate for IM and PM demodulations, respectively. Equation (\ref{ins_rate}) states that having a shorter chirp duration, $T_c$, increases the throughput. However, this may also cause more demodulation errors and affect the radar sensing since shorter chirps result in a lower signal-to-noise ratio (SNR) on target returns as a result of lower integration gain.

\section{Sensing Performance Metrics}

The sensing performance is evaluated in simulations with three performance metrics: ambiguity function (AF), Cramér–Rao lower bound (CRLB), and integrated sidelobe level. 

\subsection {Ambiguity Function}\label{AF_section}

For a radar waveform defined by $\mathbf{x}\left(t\right)$, the ambiguity function given by \cite{Abatzoglou1998},
\begin{equation}
AF\left(\tau,\omega\right)=\left|\int_{-\infty}^{+\infty}\mathbf{x}\left(t\right)\mathbf{x}^{*}\left(t-\tau\right)e^{-j2\pi\omega t}dt\right|,
\end{equation}
where $\tau$ and $\omega$ denote delay and Doppler frequency, respectively.
The ambiguity function of a waveform gives insight into the estimation of target range and velocity since it presents the sidelobes of the waveform on the delay and Doppler domains.

\begin{figure*}
    \centering
\includegraphics[width=0.7\linewidth]{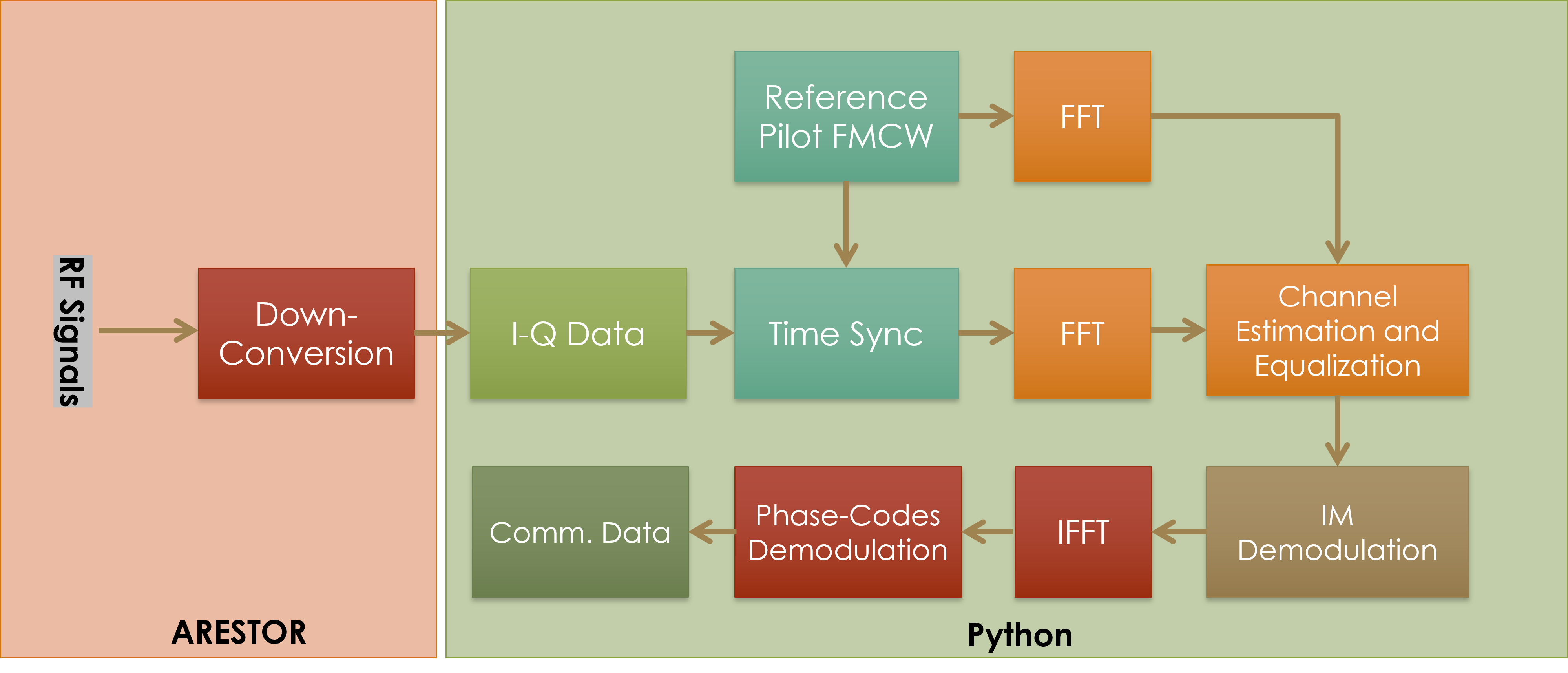}
    \caption{Communication signal processing and demodulation for each polarization in the receiver.}
    \label{fig:com_processing}
\end{figure*}

\subsection{Cramér–Rao Lower Bound (CRLB)}\label{CRLB_section}

CRLB on range indicates the theoretical lower bound of range estimation errors in radar systems, and it is frequently used to evaluate the performance of sensing systems. {The CRLB on the range is mainly dependent on the waveform bandwidth, waveform type, SNR, and chirp duration, making it a suitable radar performance metric for this study since the chirp bandwidth and center frequency vary from chirp to chirp as a result of modulating the chirps via IM. An analytical expression to calculate CRLB on the range for a complex-valued FMCW chirp with bandwidth $b_i$ can be given by \cite{ScherrFMCW2017},
\begin{equation}
    \mathrm{CRLB}_{r}
    = \frac{3c_0^2 N_s}{8\pi^2 \psi\, (N_s^2-1)\, b_i^2} \approx
    \frac{3c_0^2}{8\pi^2 \psi\, N_s\, b_i^2},
\end{equation}
where $c_0$, $\psi$, $b_i$, and $N_s$ denote the speed of light, the signal-to-noise ratio (SNR),  chirp bandwidth, and samples per chirp, respectively.}

{For IM-PM-FMCW, the same approach applies, except that the chirp bandwidth varies due to the IM. For a set of $N_b$ candidate bandwidths, the average range-estimation CRLB is given by
\begin{equation}
\overline{\mathrm{CRLB}}_r
    =
    \frac{1}{N_b}
    \sum_{i=1}^{N_b}
    \frac{3c_0^2}{8\pi^2 \psi\, N_s\, b_i^2}.
\end{equation}
This expression captures the bandwidth-dependent contribution to the CRLB for IM-PM-FMCW and provides a useful benchmark for comparison with the corresponding range-estimation performance. {The root CRLB (RCRLB) is defined as $\text{RCRLB}_r = \sqrt{\text{CRLB}_r}$ to enable a direct, unit-matched comparison with the root mean square error (RMSE) of the target range estimation.} }

\subsection{Integrated Sidelobe Level}\label{ISL_section}

Although the CRLB on the range equation is a suitable performance metric to evaluate the sensing performance of the index-modulated FMCW, it does not take into account the PM that arbitrarily alters the phase of the FMCW signals. Accordingly, we have evaluated the ratio of the main lobe to the integrated side lobe level (ISL) as another sensing performance metric \cite{UtkuPCFMCW2023}. The ISL compares the main lobe of the ambiguity function to the sum of the sidelobes within an interval. The ISL is defined as,
\begin{equation}
    \text{ISL} = 10\log10 \left[\frac{\int_{z_0}^{z_1} AF(\tau,0)d\tau+\int_{z_2}^{z_3}AF(\tau,0)d\tau}{\int_{z_1}^{z_2}AF(\tau,0)d\tau}\right], 
\end{equation}
where $AF(\tau, 0)$ denotes the ambiguity function of the signal at zero Doppler cut. Moreover, $z_1$ and $z_2$ denote the lower and upper points of the mainlobe of the ambiguity function, while $z_0$ and $z_3$ denote the bounds of the ambiguity function.

{\subsection{Fractional Out-of-Band Emission}
The out-of-band (OOB) emission is quantified using the fractional OOB power of the transmitted baseband waveform
$x(t)$. Let $S_x(f)$ denote the two-sided power spectral density (PSD) of $x(t)$ and recall $b_i$ denotes the 
bandwidth of the $i$th chirp, such that the nominal occupied band is $|f|\le b_i/2$. The average OOB in dB is given by $\Lambda = \mathbb{E}[\Lambda_i]$, and $\Lambda_i$ is given by
\begin{equation}
\Lambda_i =\; 10\log_{10}\!\left(\frac{\int_{|f|>b_i/2} S_x(f)\,df}
{\int_{-\infty}^{\infty} S_x(f)\,df}\right),
\end{equation} where the numerator represents the total radiated power outside the band, while the denominator is the total transmitted power within a chirp.}

\section{Hardware Implementation}

{A proof-of-concept hardware implementation of the proposed radar-centric ISAC system is performed based on UCL's ARESTOR platform.} The ARESTOR is a reconfigurable, multi-role RF sensor platform developed on an AMD Xilinx Radio Frequency System on Chip (RFSoC) \cite{FarleyRFSoC2018, allan2023software}. The RFSoC chip has a Field Programmable Gate Array (FPGA), an ARM central processing unit (CPU), and high-speed analog-to-digital converters (ADCs) and digital-to-analog converters (DACs) up to 6 GSPS. The ARESTOR platform offers a versatile RF and signal processing platform that can function in several sensing or communication modes \cite{Peters2021Arestor}. For instance, ARESTOR can support FMCW active radar in addition to a passive radar implementation in various frequency bands \cite{Ritchie_2022} or implementation of ISAC systems \cite{temiz_2023}. 

In the experimental measurements, the ARESTOR node is used as an ISAC transmitter, radar receiver, and communication receiver. The Tx output is fed to a 2-way power splitter; one splitter output connects to a 22-meter loopback cable, which feeds the communication receiver ADC. The second splitter output is routed through a secondary RFSoC, which introduces a variable time delay to emulate a moving target. This modified signal is then fed back to the ARESTOR radar receiver. The hardware setup is shown in Fig.~\ref{fig:exp_real_time}. Thus, the expected target range is 11 meters, while the communication range is 22 meters over the cable.

The IM-PM-FMCW baseband chirps, carrying random data, are generated in Python on a computer. The generated chirps are formatted as signed 16-bit and saved as Numpy binary. These files were transferred to ARESTOR, which mixed the baseband chirps with the carrier signal (2.4 GHz) and transmitted them over the loopback cable. The ARESTOR successfully transmitted and received the ISAC waveform over the loopback cable.

Received chirps are processed in ARESTOR for sensing. IQ-received baseband signals are also recorded as Numpy binary files, formatted as signed 16-bit, for further processing to evaluate the performance. Received RF signals by the ARESTOR are processed for sensing via ARESTOR`s FMCW processing chain, such that down-conversion, deramp, and decimation are performed for each received chirp.  After taking the variable-size FFT (range correction) of the received data, the beat frequencies are obtained, corresponding to the range information of the targets. 
 
Fig.~\ref{fig:com_processing} illustrates the block diagram of the communication data processing and demodulation of IM and PM, respectively. For communication processing, the received RF signals are quantized by high-speed ADCs of ARESTOR and then down-mixed to the baseband signals. After obtaining the baseband IQ data, it is saved in ARESTOR's storage as a Numpy binary file. This file is then processed in Python in the ARM CPU of ARESTOR for real-time processing for data demodulation as shown in Fig.~\ref{fig:exp_real_time}.  Moreover, these Numpy binary files containing sensing and communication data are added with various levels of additive white Gaussian noise (AWGN) and processed in Python on a computer to evaluate sensing and communication performance and obtain the performance figures presented.

\begin{figure}
    \centering
    \includegraphics[width=0.95\linewidth]{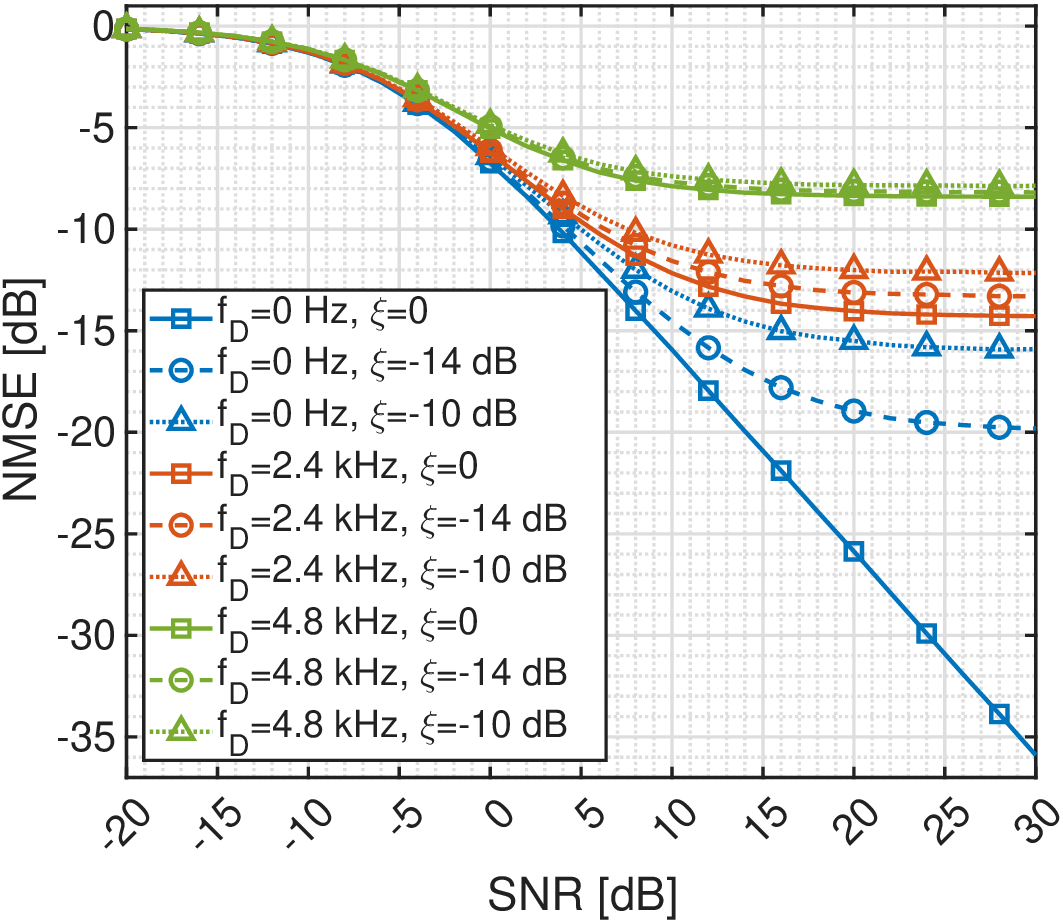}
    \caption{{The NMSE of the channel estimation under Doppler shift ($f_D$) and cross-polarisation interference ($\xi$), $T_c=50\mu$s, $b_i=50$ MHz.}}
    \label{fig:ce_results}
\end{figure}

\begin{figure}
\centering
    \includegraphics[width=1\linewidth]{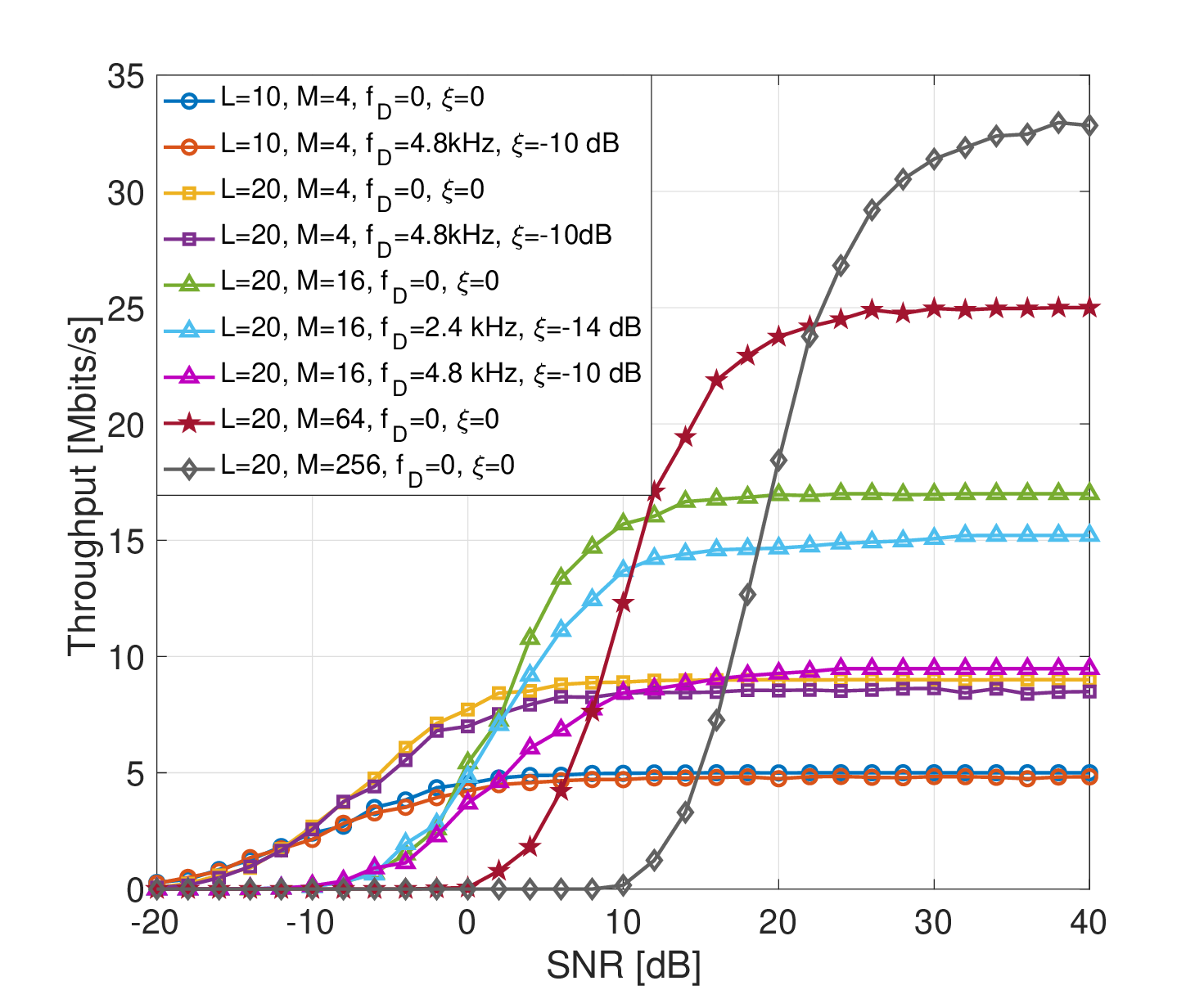}
\caption{{Throughput of the IM-PM-FMCW ISAC system under Doppler shift and cross-polarization interference, where $T_c=10 \mu s$, $\Delta f = 2$MHz,  $\Delta b = 2$MHz, $B_1$ band.}}
\label{fig:ce_rate_10us}
\end{figure}

\begin{figure}
    \centering
    \includegraphics[width=0.9\linewidth]{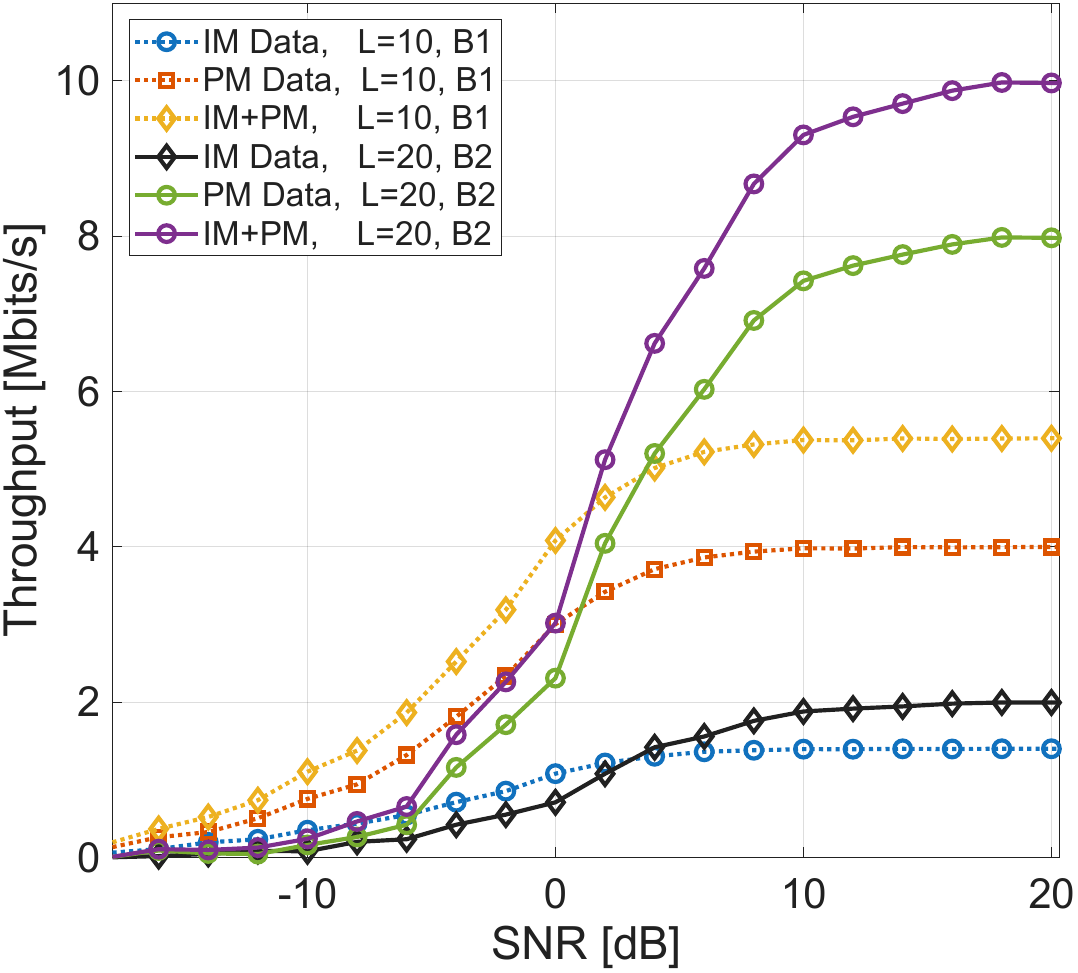}
    \caption{{Throughput achieved via IM and PM using $B_1=\{40,55\}$ MHz at $F_c=2.4$ GHz and  $B_2=\{150,250\}$ MHz at $F_c=24$ GHz bands, $T_c=10\mu$s, $M=4$.}}
    \label{fig:IM_PM_data}
\end{figure}

\begin{figure}
    \centering
    \includegraphics[width=0.90\linewidth]{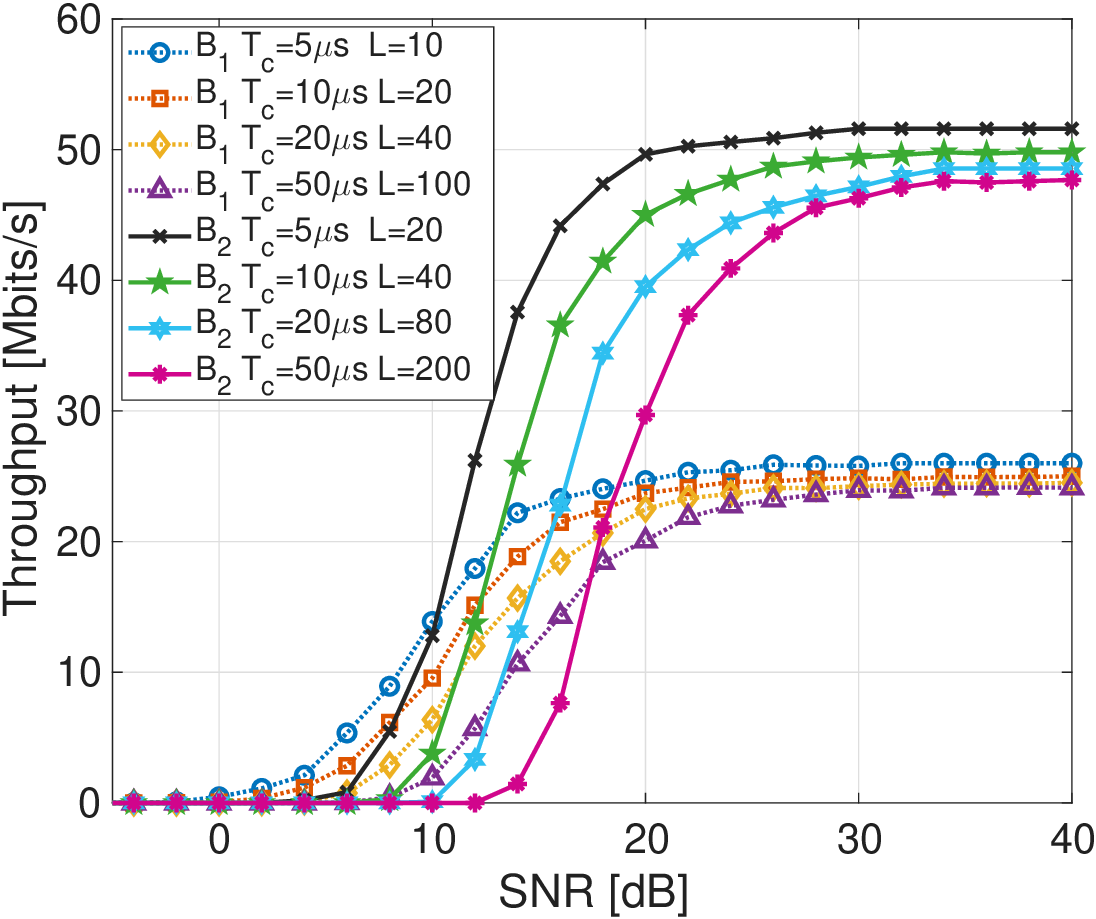}
    \caption{{Throughput achieved using $B_1=\{40,55\}$ MHz at $F_c=2.4$ GHz and  $B_2=\{150,250\}$ MHz at $F_c=24$ GHz bands, $M=64$.}}
    \label{fig:IM_PM_data_highrates}
\end{figure}

\begin{figure}
    \centering
    \includegraphics[width=0.95\linewidth]{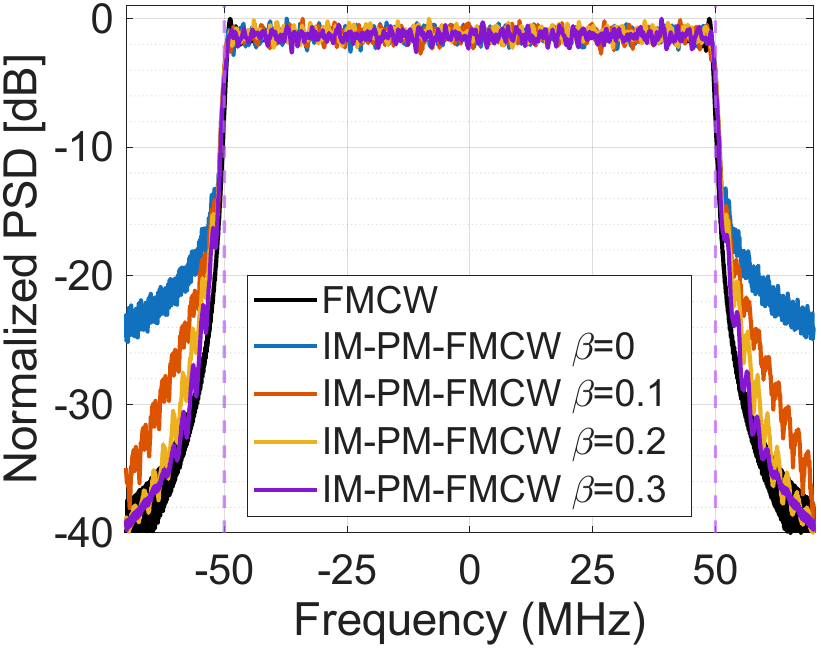}
    \caption{{Normalized power spectrum density of IM-PM-FMCW chirps with phase smoothing filter, $M=4$, $L=100$, $T_c=50\space\mu$s, $B_1$ band.}}
    \label{fig:oob_results}
\end{figure}

\begin{figure}
    \centering
    \includegraphics[width=0.90\linewidth]{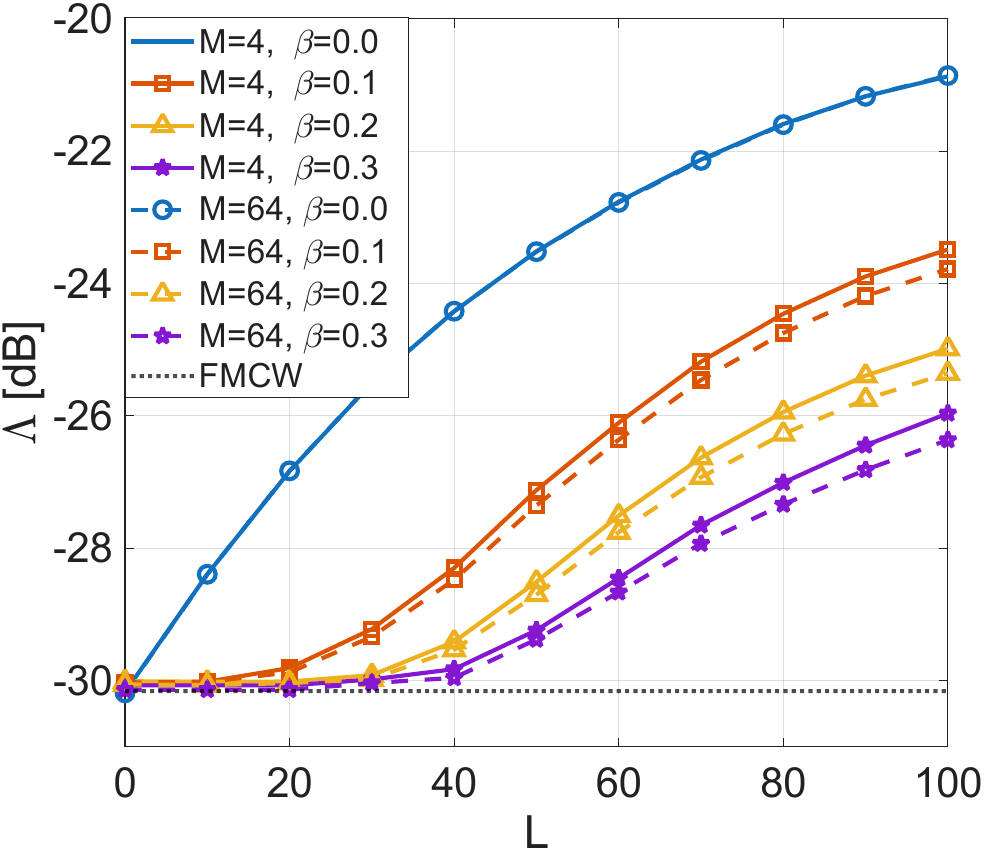}
    \caption{{Out-of-band leakage ratio ($\Lambda$) as a function of number of phase transitions ($L$), $T_c=50\space\mu$s, $b_i=50$ MHz}}
    \label{fig:oob_results_segments}
\end{figure}

In the communication receiver algorithm implemented on ARESTOR, as shown in Fig.~\ref{fig:com_processing}, firstly, time synchronization is performed to align the receiver timing with the transmitter before performing channel estimation and demodulation. The transmitted pilot FMCW chirp is also utilized for time synchronization in addition to the channel estimation. The cross-correlation of the reference pilot FMCW with the received signal over a sliding time window is calculated, and the timing of the receiver is synchronized to the time when the maximum correlation between the reference FMCW and pilot FMCW is achieved. 

\begin{figure}
    \centering
    \includegraphics[width=0.95\linewidth]{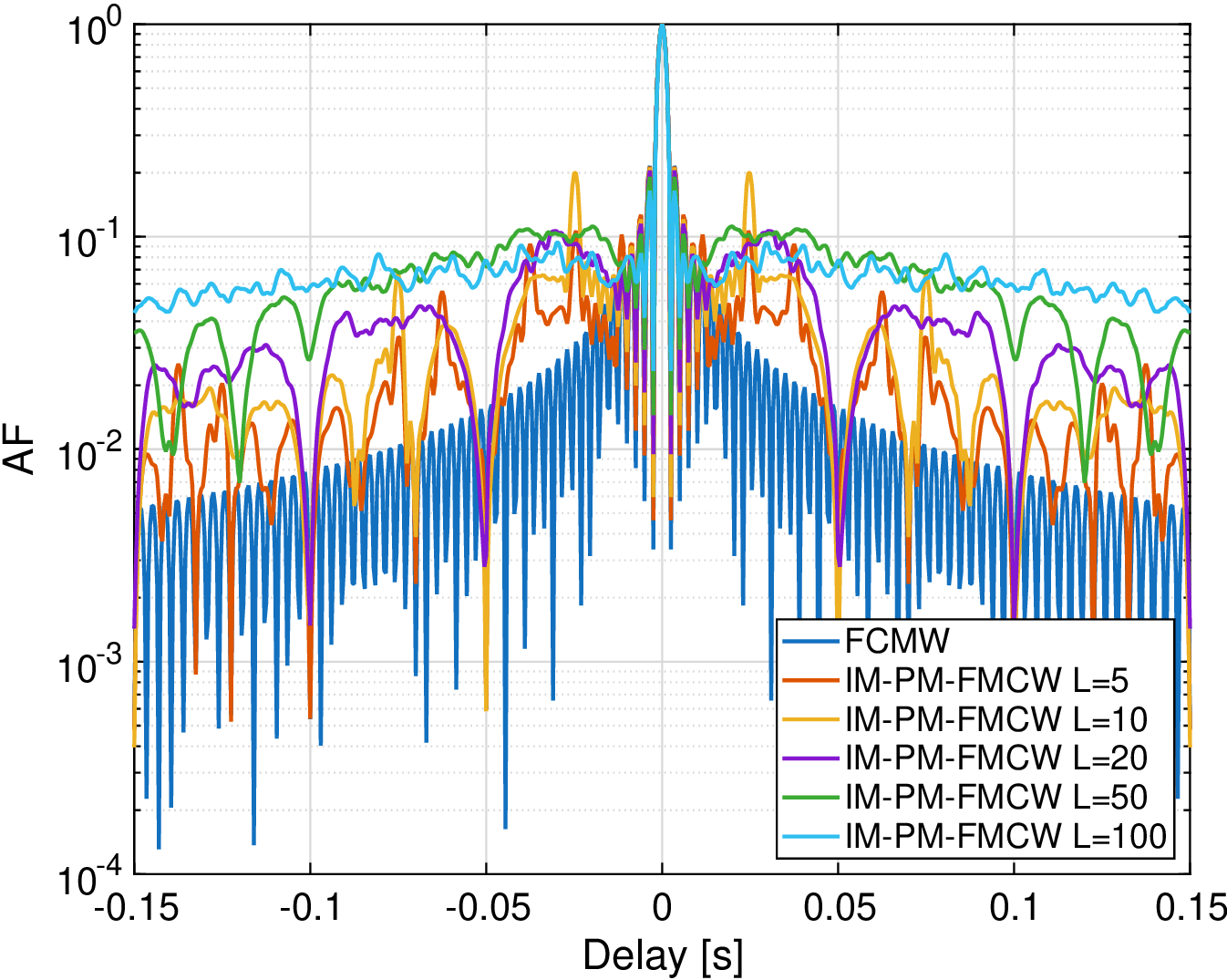}
    \caption{{Mean ambiguity function of IM-PM-FMCW signals for range estimation, $T_c=50\mu$s, $M=64$, $B_1$ band.}}
    \label{fig:AF_fig}
\end{figure}

\begin{figure}
    \centering
    \includegraphics[width=0.95\linewidth]{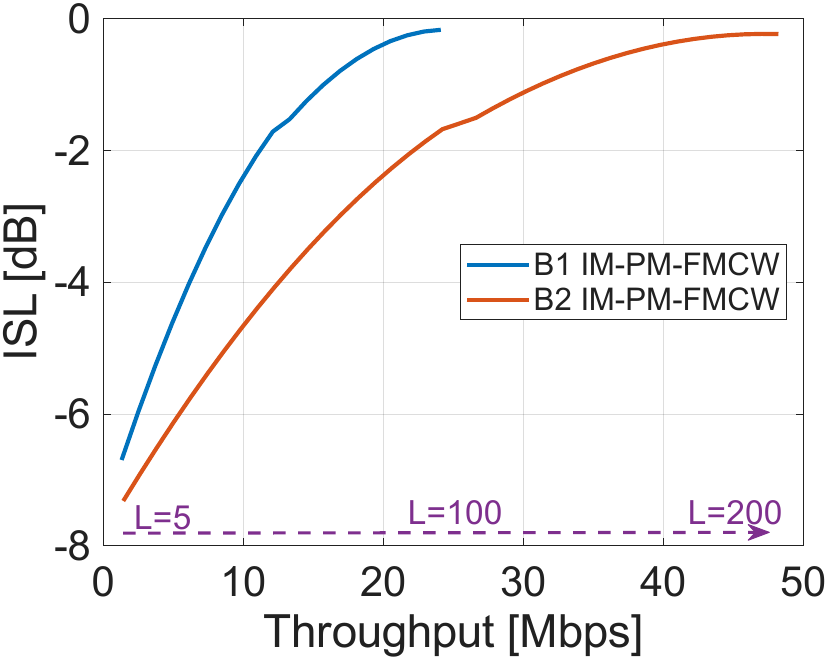}
    \caption{{The trade-off between the ISL and data rate in relation to the number of phase segments ($L$), $T_c=50\mu$s, $M=64$.}}
    \label{fig:ISL_vs_rate}
\end{figure}

After completing the time synchronization, the channel estimation and equalization are performed as shown in Fig.~\ref{fig:com_processing}. Once the channel is estimated using the pilot chirp, the subsequently received chirps in that frame are equalized in the frequency domain.  After channel estimation and equalization, demodulation of IM and PM bits is performed, as explained in subsections \ref{demod_im_section} and  \ref{Sec:Demod_PC}.

\section{Simulation and Experimental Measurement Results}

{Simulations are implemented in MATLAB and Python, leveraging the UCL ARESTOR ecosystem's native support for Python. Performance is evaluated across two ISM bands, ($B_1$: 2.4 GHz, 40–55 MHz bandwidths) and ($B_2$: 24 GHz, 150–250 MHz bandwidth). While both bands are utilized for comprehensive simulations, the physical hardware implementation is conducted exclusively in the 2.4 GHz band due to ARESTOR limitations and in accordance with UK ISM regulations \cite{interface2005uk}. Furthermore,  $I=50$ FMCW chirps are transmitted in each communication frame. Note that the proposed method does not vary the carrier frequency; instead, it varies the bandwidth and the center frequency of the baseband chirps. Therefore, we assume that the carrier frequency offset (CFO) is zero for the analysis. The results in the figure represent the average of 10,000 Monte Carlo simulation runs.}

A {Rician wideband communication channel model}, given by (\ref{channelV}), is used in simulations to investigate the performance of the channel estimation. {The Rician K-factor is configured at 3 dB with one LOS and four NLoS paths to simulate a vehicular communication scenario.} The wideband channel estimation is performed by using a single pilot FMCW chirp. In our previous work \cite{temiz_2023}, we observed that cross-polarization interference was between -14 dB and -10 dB between V-pol and H-pol in the field trials; hence, we used these values in the simulations for channel estimation. 

\subsection{Simulation Results}

\begin{figure}
\centering
  \begin{subfigure}{1\linewidth}
    \centering\includegraphics[width=0.9\linewidth]{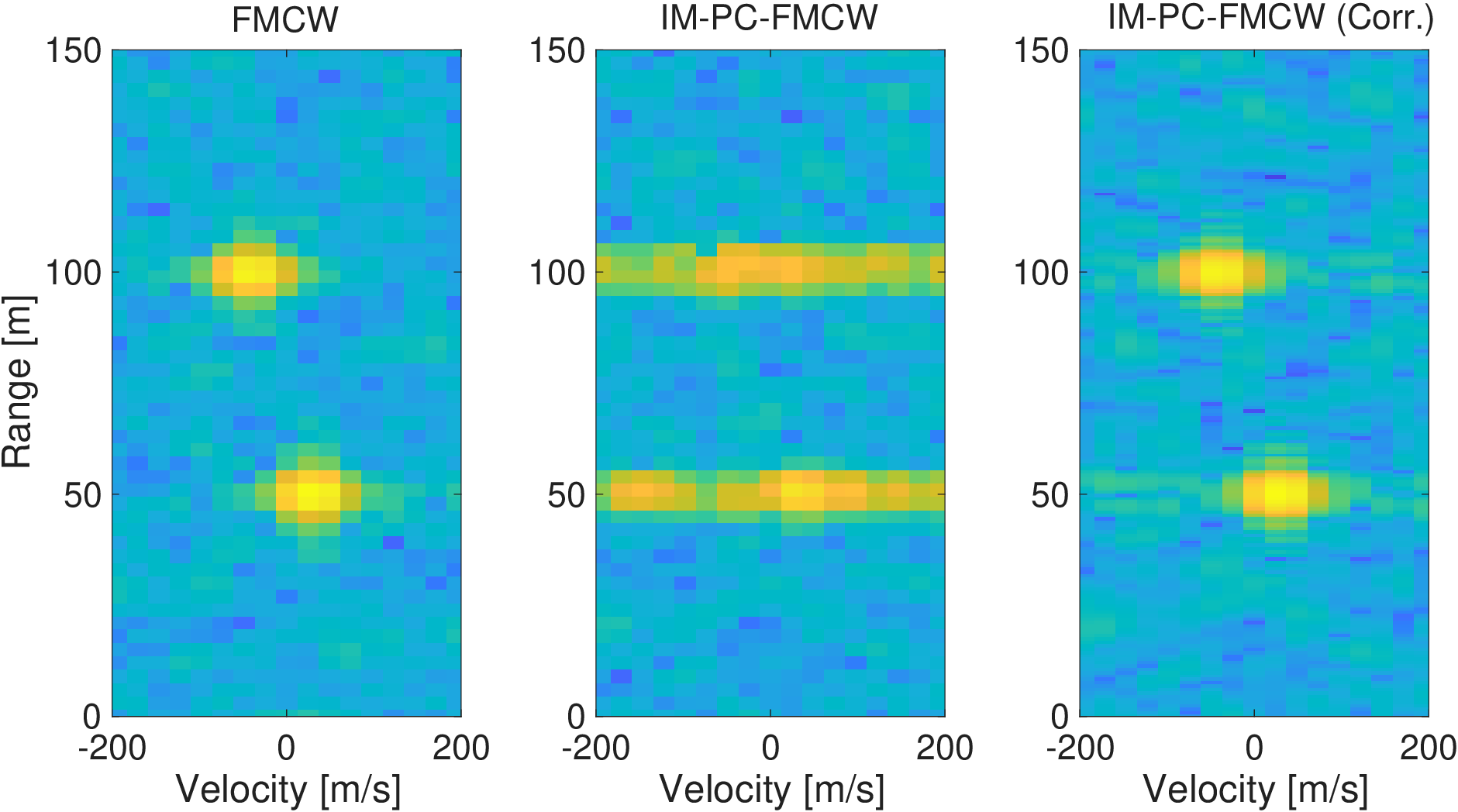}
    \caption{2.4 GHz 50 MHz.}
  \end{subfigure}%
 
  \begin{subfigure}{1\linewidth}
  \centering
\includegraphics[width=0.9\linewidth]{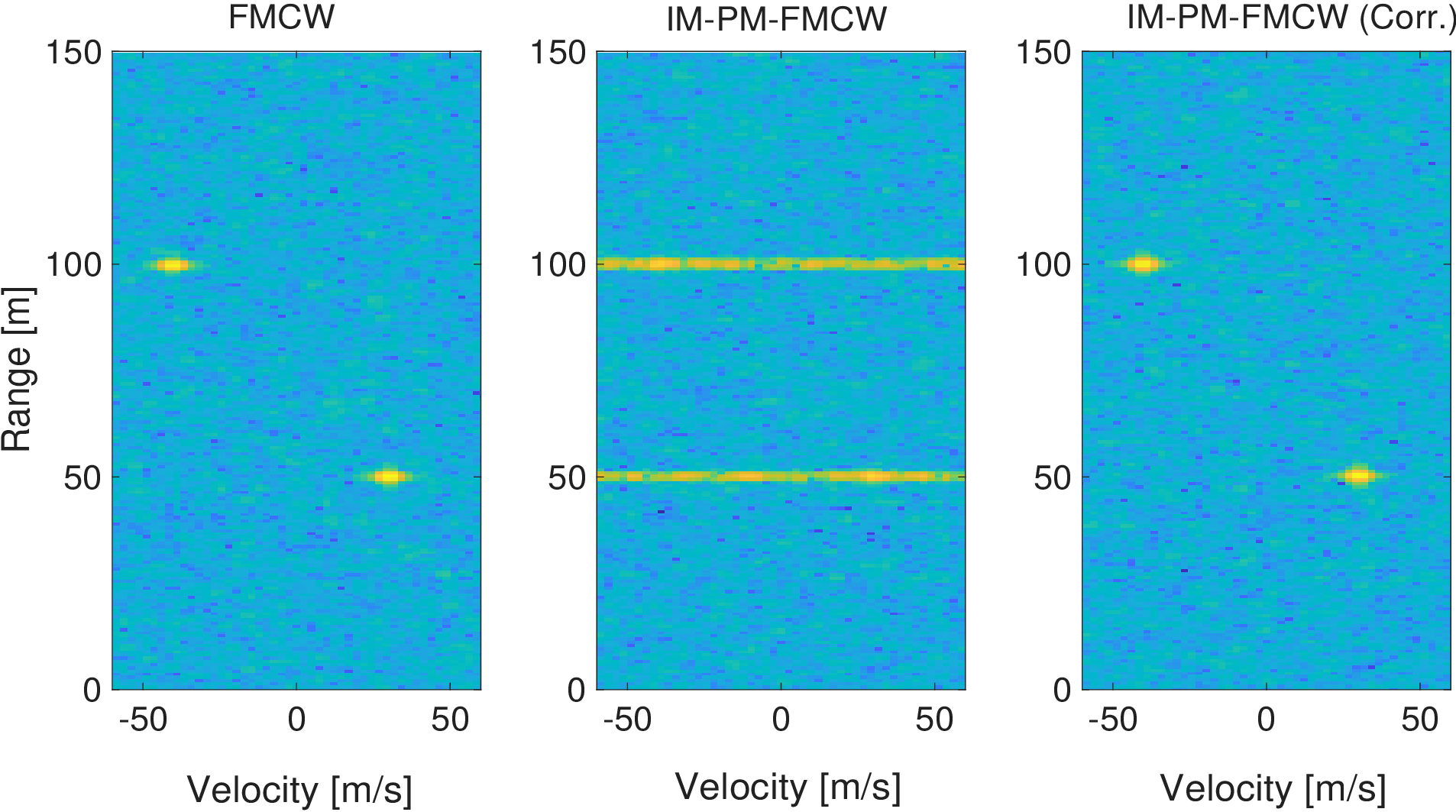}
    \caption{24 GHz 200 MHz.}
  \end{subfigure}%
\caption{{Range-velocity images of targets obtained using FMCW and IM-PM-FMCW chirps, and corrected range-velocity profile, in 2.4 GHz and 24 GHz ISM bands, $T_c=50\mu$s and $L=50$.}}
\label{fig:SimRangeVelocity}
\end{figure}

\begin{figure}
\centering
  \begin{subfigure}{.45\textwidth}
\includegraphics[width=0.85\linewidth]{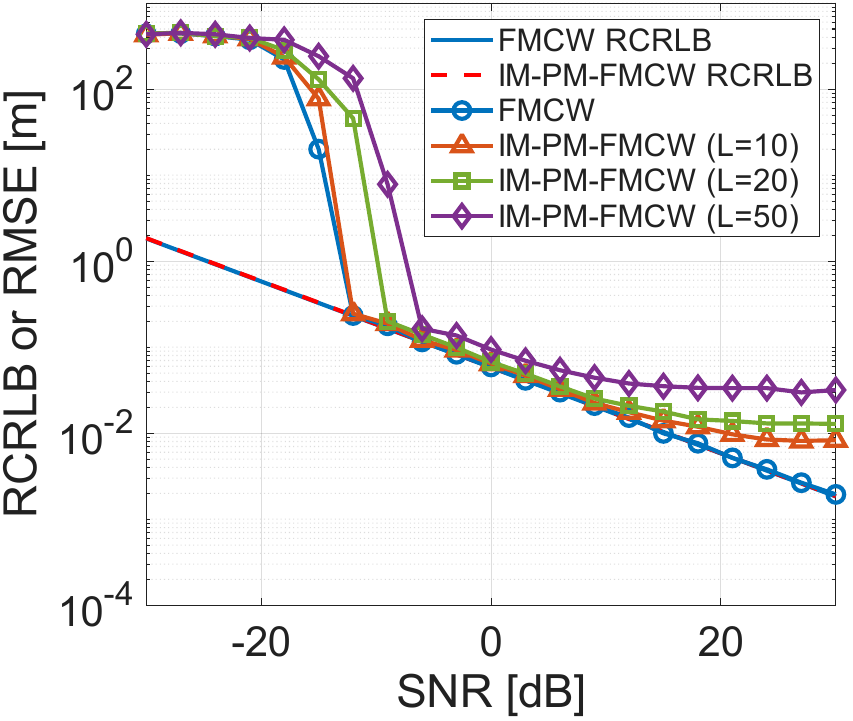}
   \centering \caption{$B_1$ band.}
  \end{subfigure}%
  \\
    \begin{subfigure}{.45\textwidth}
    \includegraphics[width=0.85 \linewidth]{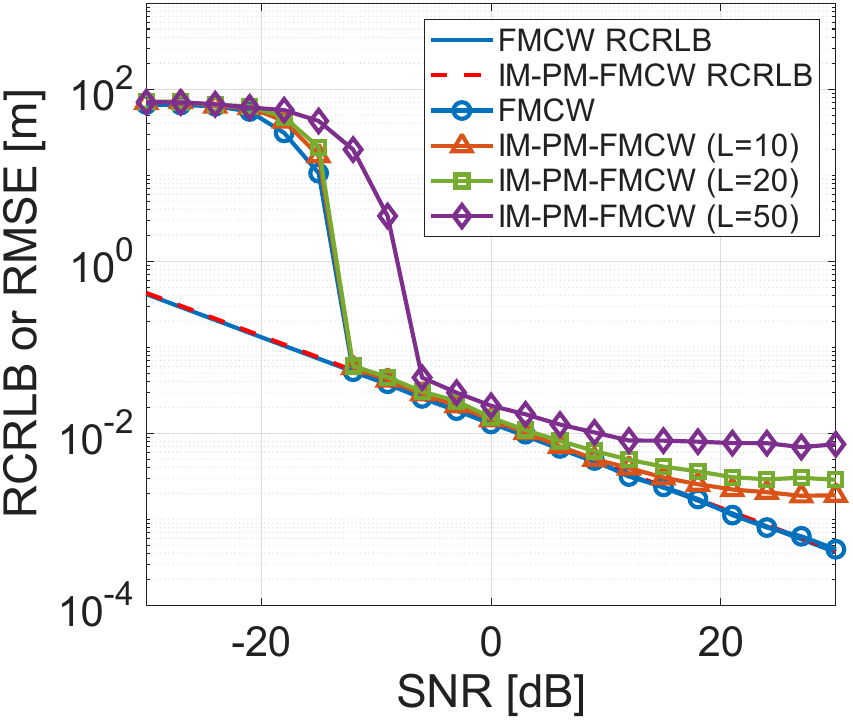}
    \centering\caption{$B_2$ band.}
  \end{subfigure}%
\caption{{Target range estimation errors and CRLB of IM-PM-FMCW in $B_1$ and $B_2$ bands, $T_c=50\mu$s.}}
\label{fig:CRLB1}
\end{figure}


Fig.~\ref{fig:ce_results} depicts the normalized mean-squared error (NMSE) of the channel estimation under the cross-polarization interference, and Doppler shift as a function of pilot SNR. Channel estimation achieves an NMSE  of -12 dB at a Doppler shift of 2400 Hz; however, performance degrades to above -8 dB as the shift increases to 4800 Hz. When operating at a 2.4 GHz carrier frequency, these shifts represent extreme vehicle velocities of 1080 km/h and 2160 km/h. Conversely, at 24 GHz, these values correspond to 108 km/h and 216 km/h, aligning with standard vehicular scenarios, respectively. Furthermore, cross-polarization interference introduces a performance floor at approximately -20 dB.

{Fig.~\ref{fig:ce_rate_10us} illustrates the data rate achieved in $B_1$ band using the proposed channel estimation method, accounting for Doppler shift, cross-polarization interference, and synchronization errors. As can be seen, reliable communication can be established with lower-order phase modulations (e.g., $M=4$ or $M=16$) under conditions of high Doppler shift and interference. Conversely, higher-order modulations can be employed in more static scenarios, achieving data rates of up to 32 Mbit/s at 2.4 GHz while utilizing approximately 50 MHz of bandwidth. Fig.~\ref{fig:ce_rate_10us} also shows that increasing the phase segments elevates the system's sensitivity to inter-symbol interference. This significantly hinders demodulation performance, particularly when subjected to Doppler shifts and cross-polarization interference.}

{Fig.~\ref{fig:IM_PM_data} illustrates the separate throughputs achieved via IM and PM, alongside the total throughput in the $B_1$ and $B_2$ bands for a given $T_c=10\mu$s and $M=4$. The results indicate that both IM and PM can reliably deliver data and be successfully demodulated under sufficient SNR conditions. Furthermore, the system achieves a higher data rate in the $B_2$ band for the same chirp duration, because the expanded bandwidth permits the transmission of more phases per chirp. Note that the bandwidth and centre frequency separations are set to $\Delta f=\Delta b=2$ MHz at $F_c=2.4$ GHz. For $F_c=24$ GHz, bandwidths of $B_2 \in \{150, 250\}$ MHz and separations of $\Delta f=\Delta b=3$ MHz are considered. Fig.~\ref{fig:IM_PM_data_highrates} presents the throughput achieved given $T_c=10$ $\mu$s, $M=64$, and an increased number of phases per chirp. These results indicate that deploying high-order phase modulations yields data rates of up to 25 Mbit/s within the $B_1$ band and 50 Mbit/s within the $B_2$ band under sufficient SNRs.}

{Conversely, the application of IM and PM within FMCW chirps impacts both sensing performance and signal power spectra. Fig.~\ref{fig:oob_results} illustrates the chirp power spectrum under phase modulation and its subsequent adjustment via the Gaussian filtering detailed in Section~\ref{sec:phase_mod}. The findings demonstrate that applying a filter with a low correction factor (e.g., $\beta=0.1$) significantly mitigates the out-of-band (OOB) emission of the IM-PM-FMCW chirps, without affecting the communication performance. Furthermore, evaluations using different phase modulation orders ($M=4$ and $M=64$) yielded similar OOB emission profiles. As depicted in Fig.~\ref{fig:oob_results_segments}, which plots OOB emission as a function of the number of phase segments ($L$) per chirp, the proposed filtering effectively suppresses OOB emissions, thereby enhancing the power spectrum of the transmitted signals. As illustrated in Fig.~\ref{fig:IM_PM_data_highrates}, while increasing $L$ allows for higher PM data rates, it also leads to increased OOB emissions, highlighting a trade-off between the two.}

{The sensing performance of the proposed IM-PM-FMCW ISAC system is further evaluated across several metrics to highlight the inherent trade-off between communication and sensing capabilities. Fig.~\ref{fig:AF_fig} depicts the ambiguity function of the IM-PM-FMCW chirps across a varying number of phase segments ($L$) per chirp. The results indicate that increasing the number of phase segments elevates the overall sidelobe levels. Consequently, a fundamental trade-off emerges between sensing and communication performance. As shown in Fig.~\ref{fig:ISL_vs_rate}, while a higher number of phase segments facilitates greater data throughput, it also leads to increased integrated sidelobe levels (ISL).}

{Fig.~\ref{fig:SimRangeVelocity} presents a comparison of the range-velocity profiles for two targets—situated at 100 m and 50 m with respective velocities of -40 m/s and 30 m/s—evaluated in both the 2.4 GHz and 24 GHz bands ($B_1$ and $B_2$). The findings indicate that the proposed IM-PM-FMCW radar receiver operates effectively, accurately compensating for range and Doppler variations as seen in Fig.~\ref{fig:SimRangeVelocity}. Furthermore, the range estimation performance is evaluated in Fig.~\ref{fig:CRLB1} for a single stationary target located 100 m away, utilizing the RCRLB as a benchmark as detailed in Section~\ref{CRLB_section}. The findings indicate that an increased number of phase segments degrades the range estimation RMSE because of elevated sidelobe levels, even though the average RCRLB of the IM-PM-FMCW chirps remains similar to that of conventional FMCW signals due to their identical average bandwidth. Additionally, as demonstrated by comparing Fig.~\ref{fig:CRLB1}(a) and Fig.~\ref{fig:CRLB1}(b), utilizing a wider bandwidth yields significantly more accurate range estimation. Moreover, the design can be scaled to other frequency bands, as it primarily performs baseband processing for communications while leveraging a standard FMCW radar receiver chain for sensing. Consequently, it does not require a higher sampling rate for the sensing function; however, it introduces additional computational complexity in the signal processing to compensate for the effects of IM and PM on sensing performance. }

\begin{figure}
\centering
  \begin{subfigure}{.45\textwidth}
  \centering
    \includegraphics[width=0.8\linewidth]{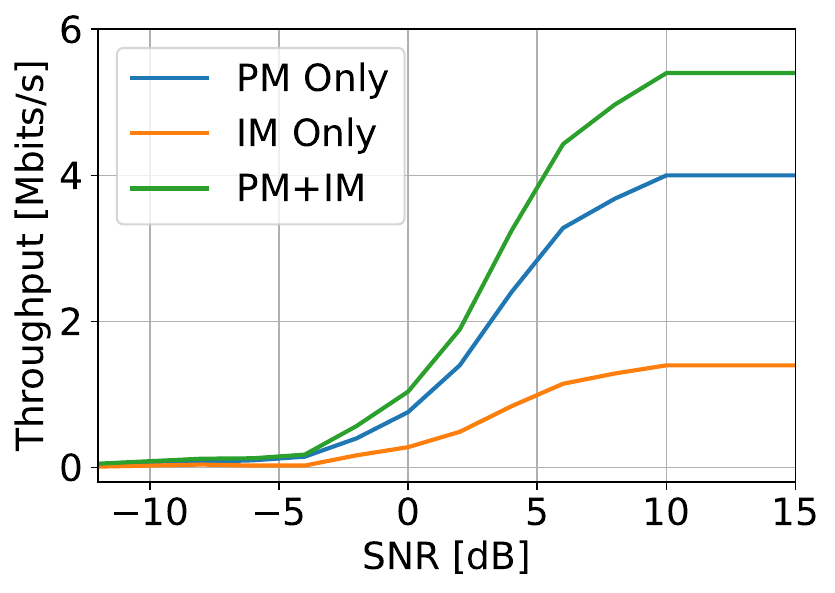}
    \caption{$T_c=10\mu s$, $L=10$ and {$\Delta f = \Delta b =2$MHz}.}
  \end{subfigure}%
  \\
    \begin{subfigure}{.45\textwidth}
    \centering
    \includegraphics[width=0.8 \linewidth]{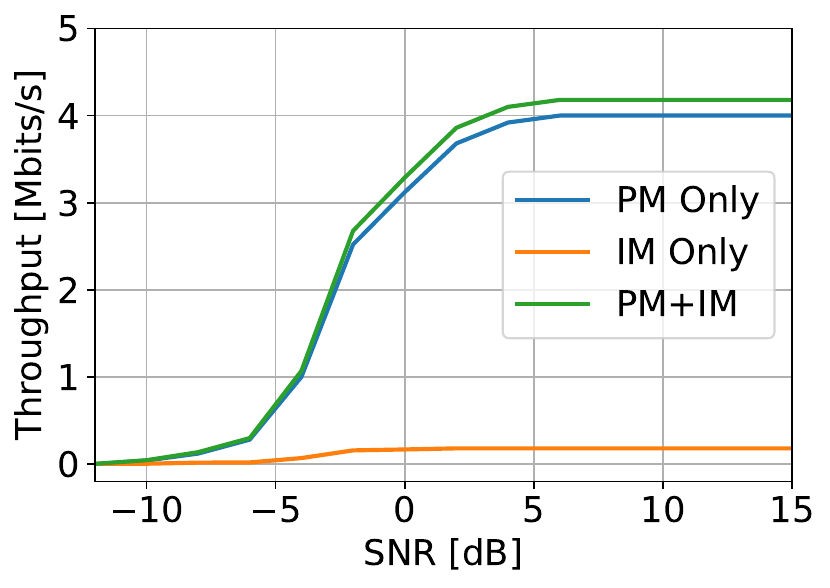}
    \caption{$T_c=100\mu s$, $L=100$, and {$\Delta f = \Delta b=1$MHz}.}
    \label{fig:rate50-100}
  \end{subfigure}%
\caption{{Throughput of IM-PM-FMCW ISAC system, $M=4$.}}
\label{fig:rate10-50}
\end{figure}


\begin{figure}
\centering
  \begin{subfigure}{1\linewidth}
    \centering\includegraphics[width=0.8\linewidth]{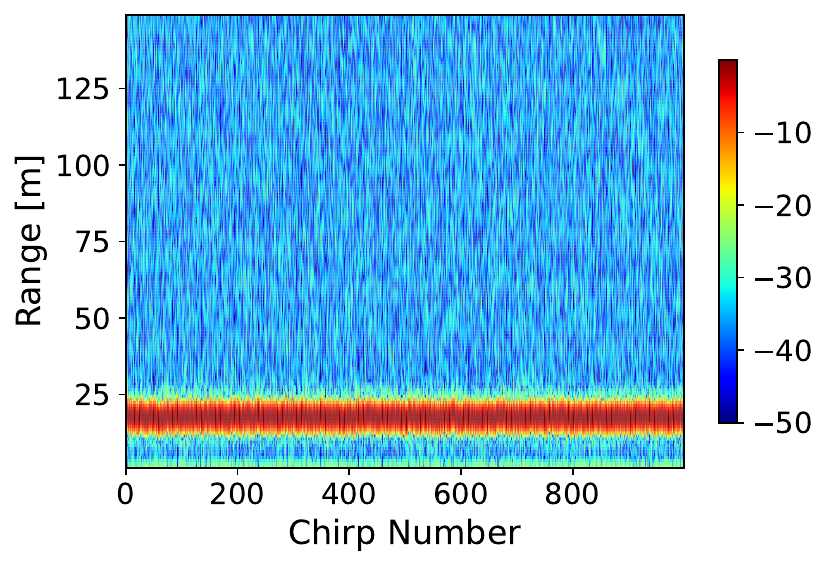}
    \caption{Before range correction.}
  \end{subfigure}%
 
  \begin{subfigure}{1\linewidth}
  \centering
\includegraphics[width=0.8\linewidth]{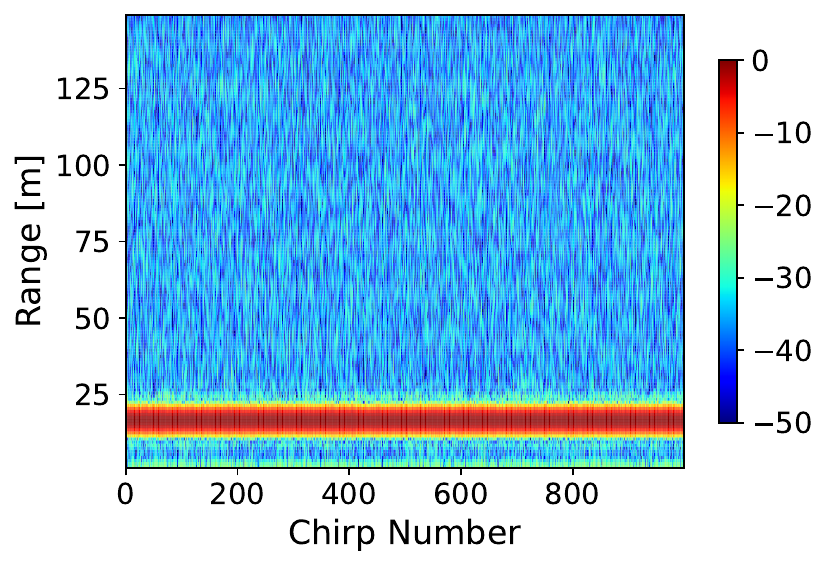}
    \caption{After range correction.}
  \end{subfigure}%
\caption{Range profile of a single target with IM-PM-FMCW signaling from experimental measurements, $T_c=50 \mu s$, $L=50$, and {$\Delta f = \Delta b =1$MHz}.}
\label{fig:IMPCFMCWSensing}
\end{figure}

\subsection{Experimental Measurement Results}

To validate the feasibility of the proposed ISAC system design, experimental measurements were conducted on the developed hardware proof-of-concept through a loopback cable, and their results and analysis are presented in this subsection. For the results presented here, the data received and processed by ARESTOR is further processed in a computer by adding the additive white Gaussian noise to obtain the SNR values presented in the figure here. Fig.~\ref{fig:rate10-50} depicts the throughput of the IM-PM-FMCW ISAC with [$T_c=10 \mu s$ and $L=10$] and  [$T_c=100 \mu s$ and $L=100$], obtained from measurements. The highest throughput is achieved when $T_c=10 \mu s$, where the IM can also significantly contribute to the final throughput since the chirp duration is short, resulting in a higher chirp repetition frequency. The throughput achieved only with PM is the same in all of these cases since the number of phase transition points is chosen in proportion to the chirp duration to make sure that a phase transition happens in each $1 \mu s$ duration. On the other hand, the throughput provided via IM proportionately decreases as the chirp duration increases since IM delivers a fixed number of bits within the chirp.  Consequently, IM and PM provide opportunities for more flexibility in ISAC waveform design, such that multiple chirp parameters can be dynamically optimized to achieve the most suitable waveform design for the desired sensing and communication trade-off. 

Sensing performance is evaluated via target range images obtained from the experimental measurement data. Fig.~\ref{fig:IMPCFMCWSensing} presents the target range profile obtained using IM-PM-FMCW chirps with  $T_c=50 \mu s$, $L=50$. Although the target is at a fixed range, the range profile in this figure varies from chirp to chirp due to varying signal bandwidth. As the bandwidth of each chirp is known by the sensing receiver, this varying range problem is mitigated by applying the proposed radar receiver technique in section~\ref{Section:Radar_processing}.

\begin{figure}
\centering
\includegraphics[width=0.8\linewidth]{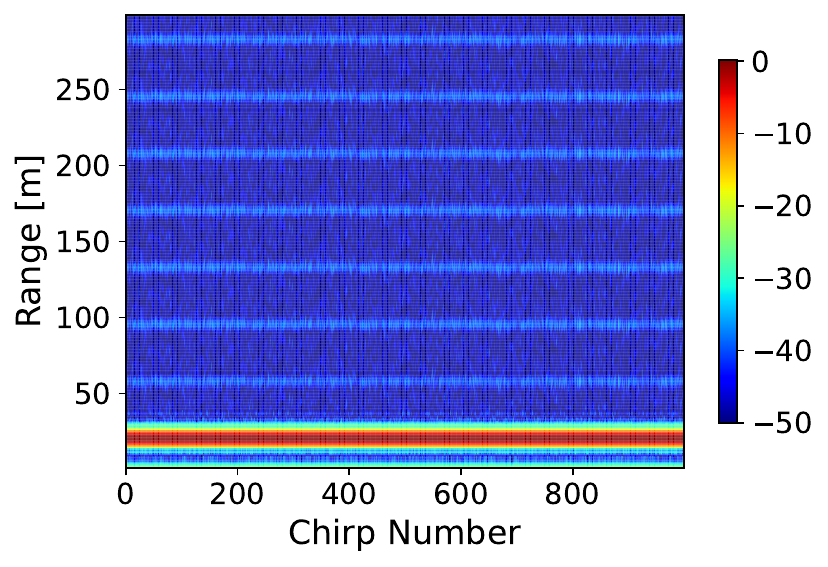}
    \centering
\caption{Range profile of a single target with IM-PM-FMCW signaling from experimental measurements, $T_c = 100 \mu s$ and $L=10$, and {$\Delta f = \Delta b =1$MHz}.}
\label{fig:IMPCFMCW_100us}
\end{figure}

To evaluate the impact of the PM further, the chirp duration is kept as $T_c=100 \mu s$ fixed while the number of phase transition points is varied as $L=10$, $L=50$, $L=100$, and the loop-back laboratory experiment is repeated with these parameters. Fig.~\ref{fig:IMPCFMCW_100us} depicts the range profiles obtained when $Tc = 100 \mu s$ and $L = 10$. It is observed that having shorter phase codes introduces fixed and repeating sidelobes; however, providing a lower average sidelobe level as seen in the comparison of Fig.~\ref{fig:IMPCFMCWSensing} and Fig.~\ref{fig:IMPCFMCW_100us}. Increasing the number of phase transition points reduced the number of strong sidelobes. However, the average sidelobe level is increased. These results are also consistent with the ambiguity function comparison of various length phase codes presented in Fig.~\ref{fig:AF_fig} and the trade-off between the communication throughput and integrated sidelobe levels shown in Fig.~\ref{fig:ISL_vs_rate}.

Experimental measurements demonstrated that the proposed IM-PM-FMCW ISAC system can deliver both communication and sensing functions using the modified FMCW chirps and can be implemented in a real-time system. Moreover, measurement results are consistent with the theory and simulation results. The proposed ISAC system provides a flexible waveform parameter design by changing the signal characteristics, such as chirp duration, number of phase transitions, or IM parameters, to adjust the trade-off between the communication throughout and sensing performance, depending on the application or scenario. {Moreover, the proposed ISAC method can be extended to multiple-input multiple-output (MIMO) systems by leveraging orthogonal phase-coded FMCW chirps to enable signal separation at the receiver. The use of orthogonal chirps would not only enable spatial multiplexing gain for communication but also create a virtual array for sensing, thereby improving angular resolution and target detection capabilities.}

\section{Conclusion}

{This study has presented a radar-centric ISAC system for automotive applications based on modified FMCW waveforms. By integrating IM and PM, the resulting IM-PM-FMCW signals successfully enable high-rate data transmission while maintaining robust sensing capabilities. A key advantage of this dual-modulation approach is the flexibility it provides in designing application-specific waveforms or dynamically adjusting parameters to balance the trade-off between communication and sensing performance. Notably, the addition of PM decouples the strict dependency between chirp duration and throughput, enabling high-speed data transfer even with longer chirp durations. The robustness of the proposed system was validated through simulations in the 2.4 GHz and 24 GHz ISM bands using two different bandwidths, achieving significant throughputs of 25 Mbps and 50 Mbps, respectively. Furthermore, a proof-of-concept implementation using loopback RF cable measurements confirms that this ISAC framework can be effectively performed on existing RF and digital processing hardware. In future work, we aim to extend our framework to account for practical hardware impairments, such as phase noise, frequency offset, and nonlinearities, to further validate the robustness of the proposed system under realistic implementation constraints.}

\section*{Acknowledgments}
This work was supported in parts by the Smart Networks and Services Joint Undertaking (SNS JU) project 6G-MUSICAL under Grant Agreement No. 101139176, and by the Engineering and Physical Sciences Research Council UK-India project ICON with Grant Agreement No UKRI859, and the AR-NGIN project of the Defence Science and Technology Laboratory (DSTL). M. Ritchie's support for this work was provided by Leonardo UK and the Royal Academy of Engineering under the Research Chairs and Senior Research Fellowships programme.

\bibliographystyle{ieeetr}
\bibliography{bibliography}

\end{document}

%% file: table1.tex
\begin{table*}[]
\centering 
\caption{Comparison of sensing-centric ISAC studies. }\label{table:literature_isac}

\begin{tabular}{|c|c|c|c|c|c|c|}
\hline
Study                                                     & Antenna                                                          & IM Indices                                                                             & Modulation                                                          & \begin{tabular}[c]{@{}c@{}}Hardware\\ Implementation\end{tabular} & Complexity & Waveform                                                     \\ \hline
\begin{tabular}[c]{@{}c@{}}MajorCom\\ (2020) \cite{HuangMajorCom2020}\end{tabular} & MIMO Array                                                       & \begin{tabular}[c]{@{}c@{}}Frequency\\ Antenna Selection\end{tabular}                  & -                                                                   & No                                                                & Moderate   & \begin{tabular}[c]{@{}c@{}}Rectangular\\ Pulses\end{tabular} \\ \hline
\begin{tabular}[c]{@{}c@{}}FRaC\\ (2021) \cite{Ma_Frac2021}\end{tabular}     & MIMO Array                                                       & \begin{tabular}[c]{@{}c@{}}Frequency\\ Antenna Selection\end{tabular}                  & Phase Mod.                                                          & No                                                                & High       & FMCW                                                         \\ \hline
\begin{tabular}[c]{@{}c@{}}FH-RCom\\ (2022) \cite{Gu_FH_2022}\end{tabular}  & MIMO Array                                                       & Frequency                                                                              & \begin{tabular}[c]{@{}c@{}}Amplitude Mod.\\ Phase Mod.\end{tabular} & No                                                                & High       & FMCW                                                         \\ \hline
\begin{tabular}[c]{@{}c@{}}DU-RCom\\ (2023) \cite{yao2023dual}\end{tabular}  & Single Antenna                                                   & Pulse Position                                                                         & Phase Mod                                                           & No                                                                & Moderate   & LFM                                                          \\ \hline
\begin{tabular}[c]{@{}c@{}}HI-RCom\\ (2023) \cite{Xu_DFRC2023}\end{tabular}  & MIMO Array                                                       & \begin{tabular}[c]{@{}c@{}}Frequency\\ Antenna\end{tabular}                            & Phase Mod.                                                          & No                                                                & High       & LFM                                                          \\ \hline
\begin{tabular}[c]{@{}c@{}}IM-FMCW\\ (2023) \cite{temiz_2023,TemizISACConf2023} \end{tabular}  & \begin{tabular}[c]{@{}c@{}}Dual-polarized\\ Antenna\end{tabular} & \begin{tabular}[c]{@{}c@{}}Frequency\\ Bandwidth\\ Polarization Selection\end{tabular} & -                                                                   & Yes                                                               & Low        & FMCW                                                         \\ \hline
This Study                                                & \begin{tabular}[c]{@{}c@{}}Dual-polarized\\ Antenna\end{tabular} & \begin{tabular}[c]{@{}c@{}}Frequency\\ Bandwidth\end{tabular}                          & Phase Codes                                                         & Yes                                                               & Moderate   & PC-FMCW                                                      \\ \hline
\end{tabular}

\end{table*}